\documentstyle[12pt]{article}

\def\proof{\noindent Proof. \hfill \break}

\def\tilde{\widetilde}
\def\bar{\overline}
\def\hat{\widehat}
\def\*{\star}
\def\({\left(}          
\def\){\right)}         
\def\[{\left[}          
\def\]{\right]}

\def\frac#1#2{{#1 \over #2}}            
\def\inv#1{{1 \over #1}}
\def\half{{1 \over 2}}
\def\d{\partial}

\def\vev#1{\langle #1 \rangle}
\def\ket#1{ | #1 \rangle}
\def\bra#1{ \langle #1 |}

\def\2pi{\hbox{$2\pi i$}}

\def\dsl{\raise.15ex\hbox{/}\kern-.57em\partial}
\def\Dsl{\,\raise.15ex\hbox{/}\mkern-.13.5mu D}
         \def\Th{\Theta}

\def\la{\lambda}        \def\La{\Lambda}
         \def\De{\Delta}

\def\vphi{\varphi}
\def\CG{{\cal G}}       \def\CH{{\cal H}}       
              
       \def\CN{{\cal N}}

\def\CY{{\cal Y}}       

\def\debut{ \begin{eqnarray} }
\def\fin{ \end{eqnarray} }
\def\non{ \nonumber }
\def\norm{:}
\def\lama{\La_{{\rm max}}}
\def\cqfd{ {\hfill{$\Box$}} }
\begin{document}
\rightline{SPhT-92-055; LPTHE-92-16}
\vskip 1cm
\centerline{\LARGE Affine  Solitons: A Relation Between}
\bigskip
\centerline{\LARGE Tau Functions, Dressing and B\"acklund
Transformations.}
\vskip 1cm
\centerline{\large Olivier Babelon }
\centerline{Laboratoire de Physique Th\'eorique et Hautes
Energies \footnote[1]{\it Laboratoire associ\'e au CNRS.}}
\centerline{
 Universit\'e Pierre et Marie Curie, Tour 16 1$^{er}$
\'etage, 4
place Jussieu}
\centerline{75252 Paris cedex 05-France}
\vskip1cm
 \centerline{\large  Denis Bernard }
 \centerline{Service de Physique Th\'eorique de Saclay
\footnote[2]{\it Laboratoire de la Direction des Sciences de la
Mati\`ere du Commisariat \`a l'Energie Atomique.}}
\centerline{F-91191, Gif-sur-Yvette, France.}
 \vskip2cm
Abstract.\\
We reconsider the construction of solitons by dressing transformations
in the sine-Gordon model.
We show that the $N$-soliton solutions are in the orbit of the vacuum,
and we identify the elements in the dressing group which allow
us to built the $N$-soliton solutions from the vacuum solution.
The dressed $\tau$-functions can be computed in two different ways~:
either using adjoint actions in the affine Lie algebra $\hat {sl_2}$,
and this gives the relation with the B\"acklund transformations,
or using the level one representations of the affine Lie
algebra $\widehat{sl_2}$, and this directly  gives the formulae
for the $\tau$-functions in terms of vertex operators.
\vfill
\newpage

\section{Introduction.}

The group of dressing transformations is the classical precursor of the
quantum group symmetry of an integrable system. It acts on the phase
space,
and its most remarkable property is that the action is
a Lie-Poisson action i.e.
to preserve the Poisson brackets, the group itself has to carry a non
trivial Poisson bracket \cite{Sem85}.
This is why after quantization the group of dressing
transformations becomes a quantum group.

Already at the classical level, this group plays a fundamental role in
our understanding of the structure of integrable two dimensionnal field
theories \cite{BaBe91,FatLu91,Lu90}.
In particular, it allows to organize
the fields into multiplets closed by dressing, and satisfying a
classical exchange algebra. In the case of a conformal field theory,
e.g. the Liouville theory, these multiplets are precisely the classical
analogues of the degenerate primary fields for which one can compute
correlation functions.

In the case of sine-Gordon, the group of dressing
transformations is infinite dimensional, and there are presumptions
that the phase space of the theory is just one orbit of the dressing
group. If this were the case, the construction of the quantum
sine-Gordon theory would just reduce to an exercise in quantum group
theory.

We will be interested in this paper to another aspect of the group of
dressing
transformations i.e. its relation to soliton solutions of the
sine-Gordon model.
 In the sine-Gordon model, particles are soliton-antisoliton bound
states, and therefore solitons are adequate to describe the asymptotic
states of the theory \cite{DAN76}.
Moreover the symmetries of the theory can be expressed conveniently
in the soliton state basis. One example is provided by the use of
local conserved charges for implementing the bootstrap program
\cite{ZaZa79,Za89}. Another important example is the quantum group
symmetry
of an integrable theory \cite{Be90,Smi91}.
These symmetries can be used for instance to determine the $S$-matrix
for the solitons.
There is some hope that correlation functions might also be determined
from this
quantum symmetry. We therefore expect a deep connection between
solitons and dressing transformations, and after quantization, between
solitons and quantum group symmetry.

There are at least two methods to generate
solitons from the vacuum solution.
One is through the Riemann-Hilbert problem
and is attached to the Zakharov-Shabat scheme
\cite{ZaSh79,NoMaPiZa84}.
The other one occurs in connexion
with $\tau$-functions and transformation groups for soliton equations
\cite{DaJiKaMi81}.
The first approach has the advantage that its Lie-Poisson properties
are now well
understood \cite{Sem85,BaBe91,FatLu91,Lu90} and this is exactly what
will become the quantum
group
symmetry of the
quantum theory. In the second approach, solitons are expressed in terms
of vertex
operators by means of remarkably simple formulae which are deeply
rooted to the algebraic structure of the theory.

The aim of this work is to establish the precise relation between these
two
descriptions of the solitons in the sine-Gordon model.

In the first approach, the dressing group is defined as an algebraic
version of the Riemann-Hilbert problem. The fields $\Phi^g$ in the
orbit of the
vacuum are given by the formula
$$e^{-2 \Lambda_\pm (\Phi^g )}= \xi_{vac}^\pm \cdot g_-^{-1} g_+ \cdot
\overline{\xi}_{vac}^\pm $$
where $\xi_{vac}^\pm$ and $\overline{\xi}_{vac}^\pm$ characterize the
vacuum, $\Lambda_\pm$ are the two fundamental highest weights of the
affine Lie algebra $\widehat{sl_2}$, and $g_-^{-1}$, $g_+$ are
triangular elements
of the dressing group.

In the second approach, the $\tau$-functions, which form
orbits
of an infinite dimensional transformation group, are defined as
expectation values
in the affine group $\hat {SL_2}$~:
$$
\tau_\pm^g(z_-,z_+)=<\Lambda_\pm|e^{-mz_+{\cal E}_+} g e^{mz_-{\cal
E}_-}|\Lambda_\pm>$$
Here, ${\cal E}_\pm$ are special elements of the affine algebra $\hat
{sl_2}$, and $g$ is an element of the transformation group.

The two formulae are in fact identical and we have
$$ \tau_\pm^g(z_-,z_+) = e^{-2 \Lambda_\pm (\Phi^g )}$$
This shows explicitly that the
transformation group of the $\tau$-function is just the group of
dressing transformations and $g= g_-^{-1} g_+$. Moreover this group is
not the affine group. Its composition law is given by
\begin{eqnarray}
h \bullet g = (h_- g_-)^{-1} h_+ g_+ \nonumber
\end{eqnarray}

We will first identify which elements of the dressing group $g_-^{-1}$,
$g_+$ generate
the $N$-soliton solutions.  They are found to be of the form~:
\begin{eqnarray}
g_-^{-1} &=& g_-^{-1}(1) \cdots g_-^{-1}(N)\non\\
g_+ &=& g_+(N) \cdots g_+(1)\nonumber
\end{eqnarray}
Each factor is given by
\begin{eqnarray}
g_-^{-1}(k) &=& e^{f_k(0) V_{\mu_k}^{(-)}} e^{{1\over 2}h_k(0) H }
e^{\half g_k(0) K} \nonumber \\
g_+(k) &=& e^{\half g_k(0) K}
e^{-{1\over 2}h_k(0) H } e^{f_k(0) V_{\mu_k}^{(+)}}\nonumber
\end{eqnarray}
where $V_{\mu_k}^{(\pm)}$, $H$, $K$, are special elements in the affine
Lie
algebra $\widehat{sl_2}$. We will then compute the $\tau$-functions in
two different ways:

\begin{itemize}
\item The first way consists in commuting $e^{-mz_+{\cal E}_+}$
and $e^{mz_-{\cal E}_-}$ through $g=g_-^{-1}g_+$ using the algebraic
structure
of $\hat {sl_2}$. In doing so, the elements $g_\pm (k)$ remain of the
same form, but $f_k$, $h_k$, $g_k$ become functions depending on
$z_\pm$. These functions are shown to satisfy differential equations
which are directly related to B\"acklund transformations.
We present an algebraic solution to them.

\item The second way consists in evaluating the $\tau$-functions by
computing the expectation values in the level one representation. We
use the level
one vertex operator representation in the principal gradation
constructed
from a single $Z_2$ twisted free boson.
The factorized elements $g=g_-^{-1}g_+$ turn out to be expressible as a
product of normal ordered
vertex operators. The result is remarkably simple and leads directly to
the
formulae of the Kyoto group.
\begin{eqnarray}
\tau_\pm^{(N)}(z_+,z_-)=\tau_0(z_+,z_-) <\Lambda^\pm \vert
\prod_{i=1}^N (1+ 2 X_i V(\mu_i)) \vert
\Lambda^\pm > \non
\end{eqnarray}
where $V_{\mu_i} =V_{\mu_i}^{(+)} + V_{\mu_i}^{(-)}$ is the vertex
operator and
  $ X_i = a_i e^{2m \left(  \mu_i z_+ + \mu_i^{-1} z_- \right) }$ are
  the
  parameters of the $N$ solitons. $\tau_0$ is the vacuum
  $\tau$-function.

\end{itemize}

We do not touch here the question of the relation between the known
Poisson
structure of the group of dressing transformations and the soliton
solutions. The hope being that when $N
\rightarrow \infty $ the $N$-soliton solutions become dense in the
orbit
of the vacuum and we would therefore obtain a convenient coordinate
system on the group of dressing transformations. We will study these
questions elsewhere.

\bigskip\bigskip
\noindent {\bf Acknowledgements:} It is a pleasure to thank L. Bonora
for his early collaboration and for his interest in this work.

\section{Lie algebra conventions.}

Let $E_+$, $E_-$, $H$ be the three generators
of the Lie algebra $sl_2$
\begin{eqnarray}
\relax [H,E_\pm ]&=& \pm 2E_\pm \nonumber \\
\relax  [E_+,E_-] &=& H \nonumber
\end{eqnarray}
We normalize the trace on $sl_2$ to $tr(HH)=2$, $tr(E_+E_-)=1$.
The loop algebra $\widetilde{sl}_2$ is the Lie algebra of traceless
$2\times 2$ matrices with entries which are  Laurent  polynomials in
$\lambda$: $ \widetilde{sl}_2= C(\lambda,\lambda^{-1})\otimes sl_2$.
The affine Lie algebra $\widehat{sl}_2$ is the central extension of
$\widetilde{sl_2}$ :
$\widehat{sl_2}=\widetilde{sl}_2 \oplus C K \oplus C d $,
with $K$ the central element and $d$ the derivation
$d=\la\frac{\d}{\d\la}$.
The bracket reads
\begin{eqnarray}
[\widehat{X},\widehat{Y}]=[\tilde{X},\tilde{Y}]_\sim +
{1 \over 2}\oint {d\lambda  \over 2i \pi} tr(\partial_\lambda
\tilde{X}(\lambda)\cdot \tilde{Y}(\lambda))\;K
\label{bracket}
\end{eqnarray}
We have the Cartan decomposition: $\widehat{sl_2}=\widehat{\cal
N}_-\oplus
\widehat{\cal H} \oplus \widehat{\cal N}_+$.
In the principal gradation we have:
\begin{eqnarray}
\widehat{\cal H}&=& \{ H,d,c\} \nonumber \\
\widehat{\cal N}_+&=&\{ E_+^{(2n-1)}=\lambda^{2n-1} E_+,
E_-^{(2n-1)}=\lambda^{2n-1} E_-, H^{(2n)}=\lambda^{2n} H, n>0 \}
\nonumber \\
\widehat{\cal N}_-&=&\{ E_+^{(2n+1)}=\lambda^{2n+1} E_+,
E_-^{(2n+1)}=\lambda^{2n+1} E_-, H^{(2n)}=\lambda^{2n} H, n<0 \}
\nonumber
\end{eqnarray}
The explicit form of the commutation relations are:
\begin{eqnarray}
\[ H^{(r)},H^{(s)} \] &=& Kr\ \delta_{r+s,0}\non\\
\[ H^{(r)},E_\pm^{(s)}\] &=& \pm2 E_\pm^{(r+s)}\non\\
\[ E_+^{(r)}, E_-^{(s)} \] &=& H^{(r+s)}+\frac{K}{2}r\delta_{r+s,0}\non
\fin
In particular, the simple roots vectors will be taken to be
$E_{\pm \alpha_1}=\lambda^{\pm 1}E_\pm $ and
$E_{\pm \alpha_2}=\lambda^{\pm 1}E_{\mp}$.
We need to define
\begin{eqnarray}
{\cal E}_+ &=& \lambda(E_+ + E_-) \label{E+}\\
{\cal E}_- &=& \lambda^{-1}(E_-+E_+) \label{E-}
\end{eqnarray}
The affine $\hat{sl_2}$ algebra possesses two fundamental highest
weights,
denoted $\La^-$ and $\La^+$. They are characterized by:
\begin{eqnarray}
\La^\pm(H)=\pm\half\quad;\quad \La^\pm(K)=1
\quad;\quad \La^\pm(d)=0\non
\fin

\section{The affine  Sine-Gordon model.}

The affine sinh-Gordon model is a Toda model over the affine
$\hat{sl_2}$ algebra.
Let $z_\pm=x\pm t$ be the light cone coordinates, $\partial_{z_\pm}=
\textstyle{1 \over 2}(\partial_x \pm \partial_t)$.
Introduce the field $\Phi$ with values in the Cartan subalgebra of
$\widehat{sl_2}$
\begin{eqnarray}
\Phi=\textstyle{1 \over 2}\varphi H + \eta d +
{1\over 4}\zeta K \nonumber
\end{eqnarray}
Following the general construction of Toda systems, we define
a Lax connexion $A_{z_\pm}$ by
\begin{eqnarray}
A_{z_+}&=&\partial_{z_+}\Phi +m e^{ad \Phi}{\cal E}_+ \label{A+} \\
A_{z_-}&=&-\partial_{z_-}\Phi +m e^{-ad \Phi}{\cal E}_- \label{A-}
\end{eqnarray}
and write the associated linear system
\begin{eqnarray}
(\partial_{z_\pm} + A_{z_\pm})T(x,t) &=& 0 \nonumber
\end{eqnarray}
where $T(x,t)$ belongs to a group whose Lie algebra is
$\widehat{sl}_2$.
The matrix $T(x,t)$ is called the transfer matrix. We normalized it
by imposing $T(0)=1$. The equations of motion
are the compatibility conditions for this linear problem. These are
the zero curvature condition
\begin{eqnarray}
F_{z_+z_-}=\partial_{z_+}A_{z_-}-\partial_{z_-}A_{z_+}+
[A_{z_+},A_{z_- }] =0 \nonumber
\end{eqnarray}
which can be worked out using only the Lie algebra structure of
$\widehat{sl}_2$. We get \cite{Man85}:
\begin{eqnarray}
\partial_{z_+}\partial_{z_-}\varphi
&=&m^2 e^{2\eta}(e^{2\varphi}-e^{ -2\varphi }) \label{Mo1} \\
\partial_{z_+}\partial_{z_-}\eta &=&0 \label{Mo2} \\
\partial_{z_+}\partial_{z_-}\zeta &=&m^2  e^{2\eta}
(e^{2\varphi}+e^{-2\varphi}) \label{Mo3}
\end{eqnarray}
Thanks to the field $\eta$, the above  equations are conformally
invariant. Moreover they are also  Hamiltonian.

In this paper, we will only be interested in the $\eta =0$ sector.
There, the equations of motion of the field $\varphi$ are those of the
sinh-Gordon
model. Changing $\varphi$ to $i \varphi$ we obtain the sine-Gordon
model
but we will not discuss here these reality conditions.

As usual, to any highest weight vector $\ket{\lama}$ we associate two
sets
of fields $\xi(x,t)$ and $\bar \xi(x,t)$ defined by \cite{Ba88}~:
\begin{eqnarray}
\xi(x,t) &=& <\lama|e^{-\Phi}T(x,t) \nonumber \\
\overline{\xi}(x,t)&=& T(x,t)^{-1} e^{-\Phi}|\lama> \nonumber
\end{eqnarray}
These fields, which are the classical analogues of the quantum vertex
operators,
are chiral:
\begin{eqnarray}
\partial_{z_-} \xi =0~~~~~~~~~\partial_{z_+}\overline{\xi}=0
\nonumber
\end{eqnarray}
The Toda field can be reconstructed from them;
the reconstruction formula reads
\begin{eqnarray}
e^{-2\lama(\Phi)}=\xi(z_+) \cdot \overline{\xi}(z_-) \nonumber
\end{eqnarray}
The vacuum solution of the equations of motion is
\begin{eqnarray}
\varphi_{vac} =0~~~~~~~~\eta_{vac} =0~~~~~~~\zeta_{vac}
=2m^2 z_{+} z_{-}
\label{vac}
\end{eqnarray}
One can insert this solution into the linear system and compute the
vacuum
transfer matrix $T_{vac}(x,t)$. We get
\begin{eqnarray}
T_{vac}(x,t)
&=& e^{-{m^2\over 2} z_+ z_- K } e^{-m z_+{\cal
E}_+}e^{-m z_- {\cal E}_-} \nonumber \\
 &=& e^{{m^2\over 2} z_+ z_- K } e^{-m z_-{\cal
 E}_-}e^{-m z_+{\cal E}_+} \nonumber
\end{eqnarray}
In the last equation, we used
$\relax[ {\cal E}_+ ,{\cal E}_- ]= K $.
We can then compute the chiral fields $\xi(z_+)$ and ${\bar
\xi}(z_-)$.
We find~:
\begin{eqnarray}
\xi_{vac} (z_+)&=&
<\lama| e^{-m z_+ {\cal E}_+} \nonumber \\
\overline{\xi}_{vac}(z_-) &=& e^{mz_- {\cal E}_-} |\lama>
\nonumber
\end{eqnarray}
Notice that the reconstruction formula gives
\begin{eqnarray}
\xi_{vac}(z_+).  {\bar \xi}_{vac}(z_-)
=\exp\Bigl(-m^2 z_+ z_- \lama (K)\Bigr)\non
\fin
in agreement with eq.(\ref{vac}).

\section{Dressing Transformations}

\subsection{Generalities.}

Dressing transformations are special symmetries of non-linear
differential equations. They are defined as gauge transformations
acting on the Lax connexion and  preserving its form.
One of their most remarkable property is
that they induce Lie-Poisson actions on the phase space \cite{Sem85}.
As dressing symmetries have already been fully described in
\cite{Sem85,BaBe91},
we restrict ourselves in giving only the basic facts concerning them
that we will need in the following. We therefore describe only the Toda
model.

The dressing transformations are associated to a factorization
problem \cite{ZaSh79}.
For a Toda field theory over an algebra $\CG$, this problem is a
factorization problem in the group $G$ whose Lie algebra is $\CG$.
In the case of the affine sinh-Gordon model, the group $G$
is an infinite dimensional group whose Lie algebra is the affine
algebra ${\widehat{sl_2}}$; namely $G\equiv {\widehat{SL_2}}\equiv
(\exp\hat \CN_-)
(\exp\hat \CH)(\exp\hat \CN_+)$. The factorization problem needed to
define
the dressing transformations is as follows.
Any element $g\in {\widehat{SL}_2}$ is decomposed as~:
\begin{eqnarray}
g=g_-^{-1}\ g_+ \quad{\rm with}\quad g_{\pm}\in B_{\pm}
=(\exp\hat \CH)(\exp\hat \CN_\pm)\non
\fin
and moreover we require that $g_-$ and $g_+$ have inverse components on
the Cartan torus.
In practice it is given by half splitting the Gaussian decomposition of
$g$.
The infinitesimal version of this factorization problem consists
in decomposing any element $X\in\CG={\widehat{sl}_2}$ as
\begin{eqnarray}
X=X_+-X_- \quad {\rm with}\quad
X_\pm\in(\hat \CH \oplus\hat \CN_\pm) \non
\fin
such that $X_+$ and $X_-$ have opposite components on $\hat \CH$.

For any element $g\in {\widehat{SL}_2}$ with factorization
$g=g_-^{-1}g_+$,
the dressing transformations are defined by the following gauge
transformation~:
\begin{eqnarray}
T(x)\to T^g(x)=\Th_\pm(x)\ T(x)\ g_\pm^{-1}\label{EGi}
\fin
where $\Th_\pm(x)$ are given by factorizing in ${\hat {SL_2}}$ the
element
$T(x)gT^{-1}(x)$:
\begin{eqnarray}
\Th_-^{-1}(x)\ \Th_+(x)=T(x)\ g\ T^{-1}(x)\label{factorization}
\fin
Notice that the gauge transformation (\ref{EGi}) can be implemented
either
using $\Th_-$ and $g_-$ or using $\Th_+$ and $g_+$: the result is the
same
thanks to eq.(\ref{factorization}).
Also these transformations preserve the normalization condition
$T(0)=1$.
They induce gauge transformations on the
Lax connexion : $A_\mu = -(\d_\mu T)\ T^{-1}$ is transformed into
$A^g_\mu =- (\d_\mu T^g)\ T^g\,^{-1}$ with
\begin{eqnarray}
A^g_\mu\ =\ \Th_\pm A_\mu \Th^{-1}_\pm -
\d_\mu\Th_\pm\,\Th^{-1}_\pm\non
\fin
The factorization problem described above is devised precisely in order
that the
form of the Lax connexion is preserved by these transformations.
The proof of this statement was given in \cite{BaBe91} . It relies on
the fact
that
the gauge transformations can be implemented using either $\Th_-$ or
$\Th_+$.
One first shows that the degrees of the components of the Toda
Lax connexion are preserved by the dressing and then, one
verifies that the connexion can be written as in
eqs.(\ref{A+},\ref{A-}).

Because the form of the Lax connexion is preserved by these
transformations,
the gauge transformations (\ref{EGi}) induce
transformations of the Toda fields~:
\begin{eqnarray}
\Phi(x,t) \to \Phi^g(x,t) \non
\fin
Moreover, since the equations of motion are equivalent to the zero
curvature condition and because gauge transformations preserve this
condition, the dressing transformations are symmetries of the
equations of motion. In other words, a dressing transformation
maps a solution of the Toda equations $\Phi(x,t)$ into
another solution $\Phi^g(x,t)$ (which, in general,
possess non-trivial topological numbers).

The transformations of the Toda fields can be described as follows.
Factorize $\Th_\pm$ as :
\begin{eqnarray}
\Th_\pm(x)\ =\ K^g_\pm(x)\ M^g_\pm(x)
\nonumber
\end{eqnarray}
with $M^g_\pm\in \exp\hat \CN_\pm$ and $K^g_\pm\in \exp\hat \CH
$.
According to the factorization problem, the components of $\Th_-$ and
$\Th_+$
on the Cartan torus are inverse: $K^g_-K^g_+=1$.  Put
\begin{eqnarray}
K^g_\pm(x)\ =\ \exp(\pm \Delta^g(x) )
\quad {\rm with}\quad \Delta^g\in\hat \CH \non
\fin
Then, by looking at the exact expression of the transformed
Lax connexion $A_{\mu}^g$, one deduces that:
\begin{eqnarray}
\Phi^g(x)\ =\ \Phi(x) - \De^g(x)  \label{exiii}
\end{eqnarray}
The relation between $\Phi^g$ and $\Phi$ is non-local because $\De^g$
are expressed in a non-local way in terms of $\Phi$.  We have:
\begin{eqnarray}
\exp(\ 2\lama(\De^g(x))\ )=\bra{\lama}\ T(x)
gT^{-1}(x)\ \ket{\lama}\non
\fin
for any highest weight $\lama$.

It should be noted that the composition law in the group of dressing
transformation is not the composition law in the group
$G$
on which the Toda model is defined. Indeed, in the dressing group
the product of two elements $(g_-,g_+)$ and $(h_-,h_+)$ is~:
\begin{eqnarray}
(g_-,g_+)\cdot (h_-,h_+)\ =\ (g_-h_-,g_+h_+)\non
\fin
In particular, the plus and minus components commute.
The brackets $[X,Y]_R$ in the dressing Lie algebra follows from the
composition for infinitesimal group elements $g=g_-^{-1}g_+\sim
1+X_+-X_-$
and $h=h_-^{-1}h_+\sim 1+Y_+-Y_-$~:
\begin{eqnarray}
[X,Y]_R\ =\ [X_+,Y_+] - [X_-,Y_-] \non
\fin

As we already mentionned, the action of dressing transformations is not
a
symplectic action, but a Lie-Poisson action \cite{Sem85}.
This means that the Poisson brackets transform covariantly only if
the group of dressing transformations is equipped with a non-trivial
Poisson structure. The Poisson structure on the dressing group is
given by the Semenov-Tian-Shansky Poisson brackets .

Since the dressing transformations are not symplectic they are not
generated by Hamiltonian functions. However there exists a non-Abelian
generalization of these Hamiltonians which applies to Lie-Poisson
actions
\cite{Lu90,BaBe91}.
The non-Abelian Hamiltonian generating the
dressing transformations  is the monodromy matrix $T$. Namely,
the variation $\delta_X\Phi(x,t)$, with $X=X_+-X_-\in\widehat{sl_2}$,
of
the field $\Phi(x,t)$ under an infinitesimal dressing transformation
is given by~:
\begin{eqnarray}
\delta_X\Phi(x,t)\ =\ Tr(\ XT^{-1}\Big\{T,\Phi(x,t)\Big\}\ )
\label{Egene}
\fin
where $Tr$ denote the trace on the affine algebra $\widehat{sl}_2$ and
$\{\, ,\,\}$ the Poisson brackets on the phase space.
The relation (\ref{Egene}) is enough to prove that the dressing
transformations have a Lie-Poisson action and this was used in
\cite{BaBe91} to describe the relation between dressing and quantum
group
symmetries.

\subsection{Tau-functions and dressings}

In Toda field theories, the number of $\tau$-functions
is the rank of the Lie algebra. They are defined as the expectation
value of the exponential of the Toda field between the fundamental
highest weight vectors $\ket{\lama^{(i)}}$:
\begin{eqnarray}
\tau_i(x,t)\ =\ \bra{\lama^{(i)}}\exp\({-2\Phi}\)\ket{\lama^{(i)}}\
	=\ \xi^{(i)}(z_+)\cdot {\bar \xi}^{(i)}(z_-)
\fin
In the affine $\hat {sl_2}$ case, there are two fundamental weights,
$\La^+$ and $\La^-$, and therefore two $\tau$-functions:
$\tau_+ = \exp{-2 \Lambda^+(\Phi)}$ and
$\tau_- = \exp{-2 \Lambda^-(\Phi)}$.
Explicitly, we have,
\begin{eqnarray}
e^{-\varphi }&=& {{\tau_+}\over{\tau_-}} \nonumber \\
e^{-\zeta } &=& \tau_+ \tau_- \nonumber
\end{eqnarray}
In terms of the $\tau$-functions,
the equations of motion become Hirota bilinear equations:
\begin{eqnarray}
(D_x^2 - D_t^2)\tau_+\cdot \tau_+ &=& -8m^2 e^{2\eta} \tau_-^2
\nonumber \\
(D_x^2 - D_t^2)\tau_-\cdot \tau_- &=& -8m^2 e^{2\eta} \tau_+^2
\nonumber
\end{eqnarray}
where the Hirota operators $D_\nu$ are defined by:
\begin{eqnarray}
D^n_\nu\ (f.g)(x)\ =\ \({\frac{\d}{\d u_\nu}}\)^n
f(x+u)g(x-u)\Big\vert_{u=0} \non
\fin
When $\eta =0$ this is just the Hirota form of sinh-Gordon.
For example, the $\tau$-functions of the vacuum solution are~:
\begin{eqnarray}
\tau_+ = \tau_- =\tau_0= \exp [-m^2 z_+ z_- ]
\nonumber
\end{eqnarray}
As we proved in \cite{BaBe91}, under dressing transformations by an
element
$g=g_-^{-1}g_+$ of the dressing group, the chiral fields
$\xi_{vac}$ and ${\bar \xi}_{vac}$ become $\xi^g$ and $\bar \xi^g$
with~:
\begin{eqnarray}
\xi^g(x,t)&=&\xi_{vac}(x,t)\cdot g_-^{-1} \non \\
{\overline \xi}^g(x,t) &=& g_+\cdot {\overline \xi}_{vac}(x,t)
\label{dress}
\end{eqnarray}
Notice how the factorization problem in the group ${\widehat{SL}_2}$
is linked with the splitting of the chirality.
By dressing the vacuum $\tau$-function $\tau_0$ is transformed
into the dressed $\tau$-functions $\tau_\pm^g(x,t)$:
\begin{eqnarray}
\tau_\pm^g(x,t)\ &=&\ \bra{\La^\pm}\exp\({-2\Phi^g}\)\ket{\La^\pm}\non
\\
&=&\ \xi_{vac}^\pm(z_+)\,g_-^{-1}\cdot g_+\, {\bar
\xi}_{vac}^\pm(z_-)\non
\fin
Using the exact expression of the vacuum transfer matrix, and hence of
the
vacuum chiral fields $\xi_{vac}$ and $\bar \xi_{vac}$, we find an
alternative
expression of the dressed $\tau$-functions:
\begin{eqnarray}
\tau_\pm^g(x,t) &=& \bra{\La^\pm} e^{-mz_+{ \cal E }_+} g_-^{-1}g_+
 e^{mz_-{ \cal E }_-}\ket{\La^\pm} \nonumber \\
 &=&\tau_0 (x,t)~ \bra{\La^\pm}\
e^{-mz_-{ \cal E }_-}\ e^{-mz_+{ \cal E }_+}\ g_-^{-1}g_+\
e^{mz_+{ \cal E }_+}\ e^{mz_-{ \cal E }_-}\ \ket{\La^\pm}\label{ETau}
\fin
Hence, by acting with the dressing group we generate
new $\tau$-functions which form the orbit of the vacuum
$\tau$-function under the dressing group.
Since the dressing transformations are symmetries of the equations
of motion, all the $\tau$-functions $\tau_i^g$ are solutions
of the Hirota equations.

\subsection{The dressing problem.}

Let us spell out how the computation of solutions obtained from
dressing the vacuum solution is related to a factorization problem.
As eq.(\ref{ETau}) made it clear, the $\tau$-functions can be
written as~:
\begin{eqnarray}
\frac{\tau_\pm^g(x,t)}{\tau_0(x,t)}\ =\ \bra{\La^\pm}\
g_-^{-1}(z_+,z_-)\cdot g_+(z_+,z_-)\ \ket{\La^\pm}\label{ETaub}
\fin
where the factors $g_\pm(z_+,z_-)$ are solution of the following
factorization problem~:
\begin{eqnarray}
g_-^{-1}(z_+,z_-)\cdot g_+(z_+,z_-)\ =\
e^{-mz_-{ \cal E }_-}\ e^{-mz_+{ \cal E }_+}\ g_-^{-1}g_+\
e^{mz_+{ \cal E }_+}\ e^{mz_-{ \cal E }_-}\label{Efac}
\fin
Once again, the factorization problem (\ref{Efac}) is the one
specified in section (4.1)~: $g_\pm(z_+,z_-)$ belong to $B_\pm$
and have opposite component on the Cartan torus.
Notice that, since $\La^\pm$
 is a highest weight, only the components on the Cartan torus of
 $g_\pm(z_+,z_-)$ contribute to eq. (\ref{ETaub}) i.e once the
 factorization problem has been solved, the $\tau$-functions are known
 explicitly. In view of eq.(\ref{Efac}), what we have to do is to
 commute
 $e^{-mz_+ {\cal E}_+}$ to the right and $e^{m z_-{\cal E}_-}$ to the
 left of $g_-^{-1} g_+$.

\subsection{The Dressing Group and the sinh-Gordon Hierarchy.}

Before examining the factorization problem, let us notice that we can
embed the vacuum equations of motion $\partial_{z_+} \partial_{z_-}
\zeta_{vac} = 2m^2$ into a larger hierarchy. Introduce the variables
$z_\pm^{(r)}$ for $r$ odd, and consider the following connexion
(generalizing eqs.(\ref{A+},\ref{A-}) evaluated at $\Phi_{vac}$)
\begin{eqnarray}
A_{z_+^{(r)}}^{vac} &=&{1\over 4} \partial_{z_+^{(r)}} \zeta~ K + m
{\cal
E}_+^{(r)} \nonumber \\
A_{z_-^{(r)}}^{vac} &=&-{1\over 4} \partial_{z_-^{(r)}} \zeta~ K + m
{\cal
E}_-^{(r)} \nonumber
\end{eqnarray}
where
\begin{eqnarray}
{ \cal E }^{(r)}_\pm=\la^{\pm r}(E_++E_-),\qquad{\rm with}\quad r\ {\rm
odd}\non
\fin
Since $[{ \cal E }^{(r)}_+ , { \cal E }^{(s)}_- ] =
Kr\delta_{rs}$, the vanishing of the curvature of this connexion
reduces to
\begin{eqnarray}
 \partial_{z_+^{(r)}} \partial_{z_-^{(s)}}\zeta_{vac} = 2m^2 r
 \delta_{rs}
 \nonumber
 \end{eqnarray}
A solution is
\begin{eqnarray}
\zeta_{vac}= 2 m^2 \sum_r r~ z_+^{(r)}z_-^{(r)} \nonumber
\end{eqnarray}
In the same way as before, we can calculate $T_{vac}(
z_+^{(r)},z_-^{(r)})$
\begin{eqnarray}
T_{vac}( z_+^{(r)},z_-^{(r)})&=& e^{-{m^2\over 2} \left( \sum_r r
z_+^{(r)}z_-^{(r)} \right) K }
e^{-m \sum_r  z_+^{(r)}{\cal E}_+^{(r)} } e^{-m \sum_r  z_-^{(r)}{\cal
E}_-^{(r)} } \nonumber \\
&=&  e^{{m^2\over 2} \left( \sum_r r z_+^{(r)}z_-^{(r)} \right) K }
e^{-m \sum_r  z_-^{(r)}{\cal E}_-^{(r)} } e^{-m \sum_r  z_+^{(r)}{\cal
E}_+^{(r)} } \nonumber
\end{eqnarray}
By dressing, the vacuum connexion will become
\begin{eqnarray}
A_{z_\pm^{(r)}}=\pm\d_{z_\pm^{(r)}}\Phi + \cdots +
m e^{\pm ad\Phi}{ \cal E }_\pm^{(r)}\nonumber
\end{eqnarray}
The $\tau$-functions $\tau^g_\pm$, eqs. (\ref{ETau}), are easily
generalized
to the multi-variable $\tau$-functions depending on all the coordinates
$z_\pm^{(r)}$:
\begin{eqnarray}
\({\frac{\tau_\pm^g}{\tau_0}}\)(z^{(r)}_\pm)\ =\ \bra{\La^\pm}\
e^{-m \sum_rz^{(r)}_-{ \cal E }^{(r)}_-}\ e^{-m
\sum_rz^{(r)}_+{ \cal E }^{(r)}_+}\ g_-^{-1}g_+\
e^{m \sum_rz^{(r)}_+{ \cal E }^{(r)}_+}\ e^{m \sum_rz^{(r)}_-{ \cal E
}^{(r)}_-}\
\ket{\La^\pm}\label{ETaubis}
\end{eqnarray}
where the vacuum $\tau$-function, $\tau_0$, is given by
\begin{eqnarray}
\tau_0 (z^{(r)}) = e^{-m^2 \sum_r  r z_+^{(r)}z_-^{(r)} } \nonumber
\end{eqnarray}
The factorization problem involved in the computation of these
$\tau$-functions
is the subject of the following sections.

\section{Solitons.}

The aim of this section consists in proving that the $N$-soliton
solutions of the affine sinh-Gordon model are in the orbit of
the vacuum solution under the dressing group.

The $N$-soliton solution of the sine-Gordon model is well known. The
$\tau$-functions are given by \cite{FaXX}
\begin{eqnarray}
{\tau^{(N)}_\pm \over \tau_0} = \det (1\pm { \cal V })  \label{tau+-}
\end{eqnarray}
where ${ \cal V }$ is a $N\times N$ matrix with elements
\begin{eqnarray}
{ \cal V }_{ij}=2{\sqrt{\mu_i \mu_j} \over \mu_i + \mu_j } \sqrt{X_i
X_j}
\nonumber
\end{eqnarray}
and
\begin{eqnarray}
X_i = a_i \exp \left[ 2m(\mu_i z_+ + \mu_i^{-1} z_- ) \right]
\label{Xi}
\end{eqnarray}
The parameters $\mu_i$ are interpreted as the rapidities of the
solitons and $a_i$ are related to their positions.
Expanding the determinants provides the explicit expression:
\begin{eqnarray}
{\tau^{(N)}_\pm \over \tau_0} = 1+\sum_{p=1}^N (\pm )^p \sum_{k_1<k_2 <
\cdots< k_p}
X_{k_1} \cdots X_{k_p} \prod_{k_i < k_j} \left( {\mu_{k_i} -\mu_{k_j}
\over
\mu_{k_i} + \mu_{k_j}} \right)^2 \label{tau+-dev}
\end{eqnarray}

According to the general theory of dressing transformation, Cf.
section 4, our strategy for proving that the N-solitons can be
generated
by dressing the vacuum will be to look for gauge transformations
$g_\pm (x,t)$, respectively upper and lower triangular, both relating
$A_{z_\pm}(\Phi_{sol})$ to $A_{z_\pm}(\Phi_{vac})$. For the transfer
matrix this
means~:
\begin{eqnarray}
T_{sol}(x,t)= g_\pm(x,t) T_{vac}(x,t) g_\pm^{-1}(0)
\nonumber
\end{eqnarray}
where $T_{sol}$ and $T_{vac}$ are the transfer matrices for the
soliton and vacuum solutions.  By compatibility we must have
\begin{eqnarray}
g_-^{-1}(x,t) g_+(x,t) = T_{vac}(x,t) g_-^{-1}(0) g_+(0)
T_{vac}^{-1}(x,t)
\label{soldress}
\end{eqnarray}
i.e we solve the factorization problem.
The element of the dressing group which map the vacuum solution
into the $N$-soliton solution is then identified to be
$g_\pm=g_\pm(0)$.

\subsection{Determination of $g_\pm$ for the one-soliton solution.}

The one soliton solution to eqs(\ref{Mo1},\ref{Mo2},\ref{Mo3}) reads
\begin{eqnarray}
e^{-\varphi}=  {{1+X}\over{1-X}},~~~~~~~
\zeta =\zeta_{vac} - \log (1-X^2) \label{zeta}
\end{eqnarray}
where
\begin{eqnarray}
X = a \exp \left[ 2m(\mu z_+ + \mu^{-1} z_- ) \right]
\nonumber
\end{eqnarray}
The independent parameters are $\mu ,a$. The $\tau$-functions are
\begin{eqnarray}
\tau_\pm = \tau_0 (1\pm X) \nonumber
\end{eqnarray}
where $\tau_0$ is the vacuum $\tau$-function.
\proclaim Proposition.
There exists two gauge transformations $g_\pm (x,t)$ relating the Lax
connexion $A_{z_\pm}(\Phi_{1~sol})$ to $A_{z_\pm}(\Phi_{vac})$.
They are given by
\begin{eqnarray}
g_-^{-1}(x)&=& e^{ f(x)\ V_\mu^{(-)} }~ e^{\log(1-X^2)~\frac{K}{4} }
\nonumber \\
g_+(x)&=&  e^{\log(1-X^2)~\frac{K}{4} }~ e^{f(x)\ V_\mu^{(+)} }
\nonumber
\end{eqnarray}
where $f(x) = \log ({{1+X}\over{1-X}})$ and $V^{(\pm)}_\mu$
are special elements of the affine Lie algebra $\widehat{sl_2}$ defined
by:
\begin{eqnarray}
 V_\mu^{(\pm)}\ =\ {1\over 2}H
 +{{(\lambda/\mu)}^{\pm2}\over{1-{(\lambda/\mu)}^{\pm2}}}H
+{{(\lambda/\mu)}^{\pm1}\over{1-{(\lambda/\mu)}^{\pm2}}}(E_+-E_-)
\nonumber
\end{eqnarray}
\par
\proof
Here is the outline of the proof. In the Appendix A, we describe the
computation in the affine group. Since $g_\pm(x,t)$ are
triangular elements in the affine group, we can restrict ourselves
to the loop group once the central terms have been determined.
The gauge transformation $g_- (x,t)$ is determined by the equations
\begin{eqnarray}
g_-(x,t)\left[ \partial_{z_+} +\partial_{z_+} \Phi_{vac}
+m {\cal E}_+ \right] g_-^{-1}(x,t) &=&  \partial_{z_+}
+\partial_{z_+}\Phi_{sol} +m
e^{ad \Phi_{sol}}{\cal E}_+ \label{gauge+} \\
g_-(x,t)\left[ \partial_{z_-} -\partial_{z_-} \Phi_{vac}
+m {\cal E}_- \right] g_-^{-1}(x,t) &=&  \partial_{z_-}
-\partial_{z_-}\Phi_{sol} +m
e^{- ad \Phi_{sol}}{\cal E}_- \label{gauge-}
\end{eqnarray}
To solve this equation, we first factorize the central terms according
to eq.(\ref{exiii}):
\begin{eqnarray}
g_- (x,t) = e^{{1\over 4} (\zeta_{sol} -\zeta_{vac})K}~\widetilde{g}_-
(x,t) \nonumber
\end{eqnarray}
One can then simply
forget the central extension and work in the loop group.
The equation for $\widetilde{g}_-$ in the loop group reads
\begin{eqnarray}
\left[ \partial_{z_+} + m \lambda \pmatrix{0 & 1 \cr 1 & 0} \right]
\widetilde{g}_-^{-1} =
\widetilde{g}_-^{-1} \pmatrix{{1\over 2}\partial_{z_+} \varphi_{sol} &
m\lambda
e^{\varphi_{sol}} \cr
 m\lambda e^{-\varphi_{sol}} &- {1\over 2}\partial_{z_+} \varphi_{sol}
 } \label{g-tilde}
 \end{eqnarray}
The $N$-solitons solutions correspond to solutions
having poles at $\lambda=\pm\mu_i$ where the $\mu_i$ are the rapidities
of the solitons:
\begin{eqnarray}
\tilde{g}_-^{-1}= D^{-1} +\sum_i \({ {P_{\mu_i} \over \lambda -\mu_i }
+ {P_{-\mu_i} \over \lambda +\mu_i } }\) \nonumber
\end{eqnarray}
with $D$ the diagonal matrix:  $D=\exp(-\half\varphi_{sol}H)$ and one
should understand the above formula as an expansion at $\lambda
\rightarrow
\infty $.
Parametrizing the $P_{\pm\mu}$ as follows:
\begin{eqnarray}
P_{\pm\mu_i} = {\tau_0 \over \sqrt{\tau_+ \tau_-}}\
\pmatrix{\pm A_{\mu_i} & B_{\mu_i} \cr
	- A_{\mu_i} & \mp B_{\mu_i}}
\nonumber
\end{eqnarray}
we find
\begin{eqnarray}
{\tau_-\over \tau_0} \partial_{z_+} A_{\mu_i} -
\partial_{z_+}{\tau_-\over \tau_0} A_{\mu_i} &=&
m \mu_i ( {\tau_+\over \tau_0} B_{\mu_i} + {\tau_-\over \tau_0}
A_{\mu_i} ) \nonumber \\
{\tau_+\over \tau_0} \partial_{z_+} B_{\mu_i} -
\partial_{z_+}{\tau_+\over \tau_0} B_{\mu_i} &=&
m \mu_i ( {\tau_-\over \tau_0} A_{\mu_i} + {\tau_+\over \tau_0}
B_{\mu_i} ) \nonumber
\end{eqnarray}
These equations are of the Hirota type. For the one soliton case,
the solution is easily found: $A_\mu = B_\mu = \mu X$.
The factor $\mu$ in front of $X$ is fixed by the affine group
computation
of the Appendix A. Including the central term, we get
  \begin{eqnarray}
  g_-^{-1}(x,t)=e^{{1\over
  4}\log(1-X^2)K}\pmatrix{\sqrt{{1+X}\over{1-X}}+{2X\over{\sqrt{1-X^2}}}
  {{(\lambda/\mu)}^{-2}\over{1-{(\lambda/\mu)}^{-2}}} &
  {2X\over{\sqrt{1-X^2}}}
  {{(\lambda/\mu)}^{-1}\over{1-{(\lambda/\mu)}^{-2}}} \cr
  -{2X\over{\sqrt{1-X^2}}}
  {{(\lambda/\mu)}^{-1}\over{1-{(\lambda/\mu)}^{-2}}}&
  \sqrt{{1-X}\over{1+X}}-{2X\over{\sqrt{1-X^2}}}
  {{(\lambda/\mu)}^{-2}\over{1-{(\lambda/\mu)}^{-2}}} }
  \nonumber
  \end{eqnarray}
We can rewrite $g_-(x,t)$ as the exponential of something. We find
\begin{eqnarray}
g_-^{-1}(x,t)=e^{{1\over 4}\log(1-X^2)K}
e^{-\varphi_{sol} \left[ {1\over 2}H
+{{(\lambda/\mu)}^{-2}\over{1-{(\lambda/\mu)}^{-2}}}H
+{{(\lambda/\mu)}^{-1}\over{1-{(\lambda/\mu)}^{-2}}}(E_+-E_-)\right] }
\label{g-x}
\end{eqnarray}
where the element in the exponential is to be interpreted as an
expansion
at $\lambda = \infty$.
So, we find $g_- =g_- (0)$ by replacing $X\longrightarrow a $ in
eq.(\ref{g-x}). One can compute similarly $g_+$. \cqfd

\subsection{One-soliton solution as a dressing problem.}

We have found the gauge transformations $g_-^{-1}$ and $g_+$ relating
the one soliton solution
to the vacuum solution. We know that they are indeed solutions of
the factorization problem defining the dressing transformations, cf
eq.(\ref{soldress}).
We now reconsider this factorization problem in a more algebraic
setting. We show
that
it is implied by some very remarkable commutation relations between
the elements $V_\mu^{(\pm)}$ that
appeared in
the gauge transformations and ${\cal
E}_\pm$. They are described in Appendix B. Once the algebraic content
of this one soliton solution has been clarified,
the generalization to N-solitons is easy.

To check eq.(\ref{soldress}), we have to show that
\begin{eqnarray}
 g_-^{-1}(x,t) g_+(x,t) = e^{-mz_- {\cal E}_-} e^{-mz_+ {\cal E}_+}
 g_-^{-1}(0) g_+(0)  e^{mz_+ {\cal E}_+} e^{mz_- {\cal E}_-}
 \nonumber
 \end{eqnarray}
 This means that the dependence in $x$ and $t$ of the soliton field is
 recovered by commuting the factor $ e^{-mz_- {\cal E}_-} e^{-mz_+
 {\cal
 E}_+}$ through $ g_-^{-1}(0) g_+(0)$. Let us consider first the
 commutation with ${\cal E}_+$.
\proclaim Proposition.
We have the following relations~:
\begin{eqnarray}
e^{-m z_+{\cal E}_+}\ e^{f(0) V_\mu^{(-)}} &=& e^{f(z_+)
V_\mu^{(-)}}~e^{g(z_+) \frac{K}{2}}~
{Y_+(z_+)}
\label{elem1} \\
e^{f(0) V_\mu^{(+)}}\ e^{mz_+{\cal E}_+} &=& {Y_+}^{-1}(z_+)\ e^{f(z_+)
V_\mu^{(+)}}
\label{elem2}
\end{eqnarray}
where
\begin{eqnarray}
\tanh \left({f(z_+)\over 2}\right)= \tanh \left({f(0) \over 2} \right)
e^{2m\mu z_+};~~
{d g \over d z_+} = -2m \mu [\cosh(f)-1]\nonumber
\end{eqnarray}
and $Y_+$ is determined by the equation
\begin{eqnarray}
{d\over{dz_+}} {Y_+}\cdot {Y_+}^{-1} =-m \cosh(f){\cal E}_+
+m \mu \sinh(f)\left(1-{1\over {2\mu}}ad~{\cal E}_+\right)H
\label{elem3}
\end{eqnarray}
\par
\proof
In eq.(\ref{elem1}), the existence of the matrix $Y_+(z_+)$ is ensured
by
Gaussian decomposition of the left hand side. The remarkable fact is
that
the component on $B_-$ of the right hand side is proportional to
$V_\mu^{(-)}$.
This property relies on the special commutation relations between
$V_\mu^{(-)}$
and ${ \cal E }_+$.
First consider eq.(\ref{elem1}).
Let ${\cal F} = e^{-mz_+{\cal E}_+} e^{f(0) V_\mu^{(-)}}$. We have
\begin{eqnarray}
{d\over{dz_+}}{\cal F} =- m {\cal E}_+ {\cal F} = {d\over{dz_+}}(e^{f
V_\mu^{(-)}} e^{g \frac{K}{2}}~
Y_+ )\nonumber
\end{eqnarray}
so that
\begin{eqnarray}
-m e^{-f ad~V_\mu^{(-)}} {\cal E}_+ = ({df\over{dz_+}})V_\mu^{(-)} +
({d\over{dz_+}}
{Y_+}){Y_+}^{-1} +({dg \over dz_+})~\frac{K}{2} \nonumber
\end{eqnarray}
Calculating the left hand side with the help of eq.(\ref{three}) of
Appendix B, gives the conditions
\begin{eqnarray}
{df\over{dz_+}}&=&2m\mu~ \sinh(f) \label{f} \\
{dg\over dz_+} &=& -2m \mu [\cosh(f) -1] \nonumber \\
{d\over{dz_+}} {Y_+}\cdot {Y_+}^{-1} &=&-m \cosh(f){\cal E}_+
+m \mu \sinh(f)\left(1-{1\over {2\mu}}ad~{\cal E}_+\right)H
\nonumber
\end{eqnarray}
Next, Considering eq.(\ref{elem2}), we find
\begin{eqnarray}
Y_+ {d\over dz_+} Y_+^{-1} = -({df \over dz_+}) V_\mu^{(+)} + m e^{f
ad~
V_\mu^{(+)}} {\cal E}_+ \nonumber
\end{eqnarray}
Therefore, to prove eq.(\ref{elem2}), we have to show that this
relation
is compatible with eq.(\ref{elem3}).
Using eq.(\ref{three}) of Appendix B, we get
\begin{eqnarray}
 Y_+ {d\over dz_+} Y_+^{-1} &=& \left(  -{df \over dz_+} +2 m \mu
 \sinh(f)
 \right)  V_\mu^{(+)} \nonumber \\
 &&+m \cosh(f){\cal E}_+ -m \mu \sinh(f)\left(1-{1\over
 {2\mu}}ad~{\cal E}_+\right)H
 \nonumber
 \end{eqnarray}
The coefficient of $V_\mu^{(+)}$ vanishes if eq.(\ref{f}) is
satisfied.  Using
${d\over{dz}}{Y_+}^{-1}=- {Y_+}^{-1}\left( {d\over{dz}}{Y_+}\right)
{Y_+}^{-1}$,
we see that the equation for $Y_+$ is identical to eq.(\ref{elem3}).
\cqfd

\noindent
A similar analysis can be done for the commutation of
$e^{-m z_-{\cal E}_-}$:
\proclaim Proposition.
We have~:
\begin{eqnarray}
e^{-m z_-{\cal E}_-}\ e^{f(0) V_\mu^{(-)}} &=& e^{f(z_-)
V_\mu^{(-)}}~
{ Y_-(z_-)^{-1}}
\label{elem1-} \\
e^{f(0) V_\mu^{(+)}}\ e^{mz_-{\cal E}_-} &=& {Y_-}(z_-)\ e^{f(z_-)
V_\mu^{(+)}}~e^{g(z_-) \frac{K}{2}}~
\label{elem2-}
\end{eqnarray}
where
\begin{eqnarray}
\tanh \left({f(z_-)\over 2}\right)= \tanh \left({f(0) \over 2} \right)
e^{\frac{2m}{\mu} z_-};~~
{d g \over d z_-} = - 2m \mu^{-1} [\cosh(f)-1]\nonumber
\end{eqnarray}
and $Y_-$ is determined by the equation
\begin{eqnarray}
{d\over{dz_-}} {Y_-}\cdot {Y_-}^{-1} =-m \cosh(f){\cal E}_-
-\frac{m}{\mu} \sinh(f)\left(1-{\mu\over 2}ad~{\cal E}_-\right)H
\label{elem5}
\end{eqnarray}
\par

As a result, when we commute $e^{-mz_+{\cal E}_+}$
and $e^{-mz_-{ \cal E }_-}$ through $g_-^{-1}(0) g_+(0)$,
the matrices $Y_+$ and $Y_-$ cancel and the parameter $a$ is replaced
by
\begin{eqnarray}
a \longrightarrow X= a e^{2m(\mu z_++ \mu^{-1}z_-)} \nonumber
\end{eqnarray}
as it should be. Moreover, the central term evolves according to
 $\log (1-a^2) \longrightarrow \log (1- a^2 e^{4m(\mu
z_++ \mu^{-1} z_-)})$.
Hence, we recover the one-soliton solution from this dressing.

\subsection{N-solitons}

In the previous section, we showed that the one-soliton solution
was in the orbit of the vacuum solution under the dressing group.
We did it in two different ways: first, we proved that there exist
gauge
transformations $g_\pm$, mapping the vacuum Lax connexion $A_{vac}$
into
the one-soliton Lax connexion $A_{sol}$; then, we demonstrated that
these
same gauge transformations could be found by solving a dressing
problem.

In this section, we show that the N-soliton solutions are also in the
orbit
of the vacuum under the dressing group. We will do it by solving a
dressing problem using only the algebraic properties of the affine Lie
algebra
$\widehat{sl_2}$.
 According to the discussion of section 4, this means
that,
as we will show, there exist elements $g=g_-^{-1}g_+$ in the affine
group
$\hat {SL}_2$ such that the N-soliton solution $\Phi_{N-sol}$ is
given by~:
\begin{eqnarray}
e^{-2\La^\pm(\Phi_{N-sol}-\Phi_{vac})} &=&
\({\frac{\tau^{(N)}_\pm}{\tau_0}}\)(z_+,z_-) \label{Ensol}\non\\
&=& \bra{\La^\pm}\ g_-^{-1}(z_+,z_-)\cdot
g_+(z_+,z_-)\ \ket{\La^\pm}\non
\fin
where $g_\pm(z_+,z_-)$ are defined from the following factorization
problem~:
\begin{eqnarray}
g_-^{-1}(z_+,z_-)\cdot g_+(z_+,z_-) =
e^{-mz_-{ \cal E }_-}e^{-mz_+{ \cal E }_+} g_-^{-1}\cdot g_+
e^{mz_+{ \cal E }_+}e^{mz_-{ \cal E }_-}\label{ENi}
\fin

The elements $g_\pm$ of the affine group $\hat {SL}_2$ which connect
the N-soliton solution to the vacuum solution
are products of N factors:
\begin{eqnarray}
g_-^{-1} &=& g_-^{-1}(1) \cdots g_-^{-1}(N)\non\\
g_+ &=& g_+(N) \cdots g_+(1) \label{EIVi}
\end{eqnarray}
where each factor is given by~:
\begin{eqnarray}
g_-^{-1}(k) &=& e^{f_k(0) V_{\mu_k}^{(-)}} e^{{1\over 2}h_k(0) H }
e^{\half g_k(0) K} \label{g-k} \\
g_+(k) &=& e^{\half g_k(0) K}
e^{-{1\over 2}h_k(0) H } e^{f_k(0) V_{\mu_k}^{(+)}} \label{g+k}
\end{eqnarray}
Here we choose $h_N$, which cancels in eq.(\ref{ENi}), such that
$\sum_k
h_k =0$. This ensures that $g_+$ and $g_-$ have opposite components on
the Cartan torus.
The dressing problem is solved by the following
\proclaim Proposition.
Consider the factorization problem eq.(\ref{ENi}) where the
elements
$g_\pm$ are given by eqs.(\ref{EIVi}-\ref{g+k}).\\
{\bf (a)} Its solution $g_\pm (z_+,z_-)$ is of the form~:
\begin{eqnarray}
g_-^{-1}(z_+,z_-) &=& g_-^{-1}(1)(z_+,z_-) \cdots
g_-^{-1}(N)(z_+,z_-)\non\\
g_+(z_+,z_-) &=& g_+(N)(z_+,z_-) \cdots g_+(1)(z_+,z_-)\nonumber
\end{eqnarray}
Each factor $g_\pm(k)(z_+,z_-)$,
has the same form as the corresponding factor $g_\pm(k)$ but with
$z_+$ and $z_-$ dependent functions $f_k$, $g_k$ and $h_k$.\\
{\bf (b)} These functions satisfy:
\begin{eqnarray}
\partial_{z_+} f_k &=& 2m\mu_k \sinh~(\phi_k + f_k ) -
2m\mu_k \sinh~(\phi_k ) \nonumber\\
\partial_{z_+} g_k &=& -2m\mu_k \cosh~(\phi_k + f_k ) +
2m\mu_k \cosh~(\phi_k ) \label{recur+}\\
\partial_{z_+} h_k &=& 2m\mu_k \sinh(\phi_k + f_k)
- 2m\mu_{k+1} \sinh~(\phi_{k+1}) \non \\
\phi_{k+1} &=& \phi_k + f_k +h_k \nonumber
\end{eqnarray}
and
\begin{eqnarray}
{\partial}_{z_-} f_k &=& -2m\mu_k^{-1} \sinh~({\bar \phi}_k - f_k ) +
2m\mu_k^{-1} \sinh~({\bar \phi_k}) \nonumber\\
{\partial}_{z_-} g_k &=& -2m\mu_k^{-1} \cosh~({\bar \phi}_k - f_k ) +
2m\mu_k^{-1} \cosh~({\bar \phi_k}) \nonumber\\
{\partial}_{z_-} h_k &=& 2m\mu_k^{-1} \sinh({\bar \phi}_k - f_k)
- 2m\mu_{k+1}^{-1} \sinh~({\bar \phi}_{k+1}) \label{recur-} \\
{\bar \phi}_{k+1} &=& {\bar \phi}_k - f_k +h_k \nonumber
\end{eqnarray}
{\bf (c)} The N-soliton $\tau$-functions are
then given by~:
\begin{eqnarray}
\frac{\tau^{(N)}_\pm}{\tau_0}(z_+,z_-) =
\exp\({\pm\half \sum_k\ f_k(z_+,z_-) + \sum_k\ g_k(z_+,z_-) }\) \non
\fin
\par
\proof
The commutation of $e^{-m z_+ {\cal E}_+}$ through $g_-^{-1}$ is done
in
$N$ steps. The first step is given by eq.(\ref{elem1}). In the second
step, we have to
commute $Y_+(z_+)$ through $g_-^{-1}(2)$, etc ... The general pattern
is as
follows. There exists elements $Y_+(k,z_+)$ such that
\begin{eqnarray}
Y_+(k,z_+) e^{f_k(0) V_{\mu_k}^{(-)}} e^{{1\over 2}h_k(0) H }
 e^{\half g_k(0) K} = e^{f_k(z_+) V_{\mu_k}^{(-)}} e^{{1\over
 2}h_k(z_+) H }
 e^{\half g_k(z_+) K} Y_+(k+1 ,z_+) \nonumber
 \end{eqnarray}
 The important point is that they are determined by the simple
 equations
 (compare with eq.(\ref{elem3}))
 \begin{eqnarray}
 \partial_{z_+} Y_+(k,z_+) \cdot Y_+(k,z_+)^{-1} =-m \cosh \phi_k~
 {\cal
 E}_+
 + 2m \mu_k  \sinh \phi_k ~U_{\mu_k}^{(+)} \nonumber
 \end{eqnarray}
where the element $U_{\mu_k}^{(+)}$ is defined in the Appendix B. The
proposition is a consequence of this observation. The details are given
in Appendix C. Just as in the one-soliton case, we have~:
\begin{eqnarray}
e^{-mz_+{ \cal E }_+}\ g_-^{-1} &=& g_-^{-1}(z_+)\ Y_+(N+1;z_+) \non \\
g_+\ e^{mz_+{ \cal E }_+} &=& Y^{-1}_+(N+1;z_+)\ g_+(z_+) \non
\end{eqnarray}
Hence, the matrix $Y_+(N+1;z_+)$ cancels out when commuting
$e^{-mz_+{ \cal E }_+}$
through $g=g_-^{-1}g_+$. Similarly with $e^{-m z_- {\cal E}_-}$. \cqfd

\section{Relation with the B\"acklund transformations.}

In this section we show how the dressing problem solved in the
previous section is related to the so-called B\"acklund
transformations, See e.g. \cite{SCMc73} and references therein.
More precisely, we will map the differential equations
(\ref{recur+},\ref{recur-}) into
a set of B\"acklund transformations and use this relation to
present an algebraic solution to them.

\subsection{B\"acklund transformations.}

The B\"acklund transformations are also symmetries of non-linear
differential equations but of different kind than the dressing
transformations. For the sinh-Gordon model, the B\"acklund
transformation
can be defined as follows:
Let $\varphi$ be a solution of the sinh-Gordon model, its image
under a B\"acklund transformation with spectral parameter $\mu$ is
the field $\hat \varphi \equiv B_\mu\cdot\varphi$ implicitely defined
by
the following differential equations:
\begin{eqnarray}
\d_{z_+}(\hat \vphi +  \vphi) &=& 2m\mu \sinh ( \hat \vphi -
\vphi)\non\\
\d_{z_-}(\hat \vphi -  \vphi) &=&
\frac{2m}{\mu} \sinh  ( \hat \vphi + \vphi)\label{EVi}
\fin
If $\vphi$ solves the sinh-Gordon equation, so does
the transformed field $\hat \vphi$.
This can be seen by remarking that eqs.(\ref{EVi}) implies~:
\begin{eqnarray}
\d_{z_+} \d_{z_-}(\vphi +\hat \vphi) = 2m^2\Bigl[ \sinh (2\vphi)
+ \sinh (2\hat \vphi)\Bigr] \non
\fin

In general the relation between $\vphi$ and $\hat \vphi$ is non-local.
However, when expanding $\hat \vphi$ in formal power of either $\mu$ or
$1/\mu$ each term of the infinite series is a local function
of $\vphi$. This remark can be used to deduce an infinite set of local
conserved currents in the sinh-Gordon model. Indeed, from the defining
relations (\ref{EVi}) of the B\"acklund transformation, it follows
that the current $J_{z_+},\ J_{z_-}$ with components:
\begin{eqnarray}
J_{z_+}\ &=&\ \mu  \cosh (\vphi -\hat \vphi) \non\\
J_{z_-}\ &=&\ -\inv{\mu} \cosh (\vphi +\hat \vphi) \non
\fin
is a conserved current: $\d_{z_-}J_{z_+}+\d_{z_+} J_{z_-}=0$.
Expanding it in power series of either $\mu$ or $1/\mu$ gives two
infinite series of local conserved currents.

A remarkable property of the B\"acklund transformations is that
we have the commutative diagram (see e.g. \cite{SCMc73})~:
\begin{eqnarray}
\matrix{~&~& \vphi_1 = B_{\mu_1}.\vphi_0&~&~\cr
 & \nearrow_{\mu_1} & & \searrow^{\mu_2} & \cr
\vphi_0&~ &~&~& \vphi_3 = B_{\mu_1}.\vphi_2 = B_{\mu_2}.\vphi_1 \cr
 & \searrow^{\mu_2} & & \nearrow_{\mu_1} & \cr
 ~&~& \vphi_2 = B_{\mu_2}.\vphi_0&~&~\cr} \label{EVii}
\fin
A consequence of this diagram is that the four solutions $\vphi_0,\
\vphi_1,\ \vphi_2$ and $\vphi_3$ are linked by purely algebraic
relations:~
\begin{eqnarray}
\tanh
\({\frac{\vphi_3-\vphi_0}{2}}\)=\({\frac{\mu_1+\mu_2}{\mu_1-\mu_2}}\)\
\tanh \(\frac{\vphi_2-\vphi_1}{2}\) \label{EViib}
\fin
We called this rule the ``tangent rule".
It is proved  starting from the relation~:
\begin{eqnarray}
\d_{z_+}(\vphi_3+\vphi_1) - \d_{z_+}(\vphi_1+\vphi_0) =
\d_{z_+}(\vphi_3+\vphi_2) - \d_{z_+}(\vphi_2+\vphi_0)  \non
\fin
and similarly with $z_-$, using the defining relation (\ref{EVi}) of
the B\"acklund transformation.

By repeated uses of this rule, one can construct in a purely algebraic
manner an infinite set of solutions of the sinh-Gordon equation.

The `tangent rule' can be expressed in terms of the $\tau$-functions.
If $\tau_\pm(k)$ with $k=0,\cdots,3$, are $\tau$-functions
of the fields $\vphi_k$, then we have~:
\begin{eqnarray}
\tau_+(3)\tau_-(0) + \tau_+(0)\tau_-(3) &=&
\tau_+(1)\tau_-(2) + \tau_+(1)\tau_-(2) \label{EVv}\\
\beta_{12}\Bigr[\tau_+(3)\tau_-(0) - \tau_+(0)\tau_-(3)\Bigl] &=&
\tau_+(2)\tau_-(1) - \tau_+(1)\tau_-(2) \label{EVv'}
\fin
where, for later convenience, we put~:
\begin{eqnarray}
\beta_{12} = \frac{\mu_1-\mu_2}{\mu_1+\mu_2} \non
\fin
In this form the `tangent rule' also applies to the affine sinh-Gordon
model as well.

\subsection{From Dressing to B\"acklund transformations.}

We now map the relations (\ref{recur+},\ref{recur-}) into a set of
successive B\"acklund
transformations. First, let us introduce the following change
of variables~:
\begin{eqnarray}
f_k &=& - \vphi_k + \vphi_{k-1}\non\\
h_k &=& \rho_k - \rho_{k-1} \label{EVvi}\\
g_k &=& -\half(\zeta_k -\zeta_{k-1})\non
\fin
with the initial condition $\vphi_0=\rho_0=\zeta_0=0$.
In terms of these new functions, the N-soliton $\tau$-functions are~:
\begin{eqnarray}
\frac{\tau_\pm^{(N)}}{\tau_0} = \exp(\mp\half\vphi_N - \half
\zeta_N)\label{ETN}
\fin
In particular, the functions $\vphi_N$ and $\zeta_N$ are the N-solitons
solutions of
the affine sinh-Gordon model.

In these new variables, the relations (\ref{recur+},\ref{recur-})
involving $f_k$ and $h_k$ become~:
\begin{eqnarray}
\d_{z_+}(\vphi_k+\rho_{k-1})&=&2m\mu_{k} \sinh (\vphi_k-\rho_{k-1})
\label{backlund1}\\
\d_{z_-}(\vphi_k-\rho_{k-1})&=&\frac{2m}{\mu_{k}}\ \sinh
(\vphi_k+\rho_{k-1}) \label{backlund2}
\fin
and
\begin{eqnarray}
\d_{z_+}(\rho_k +\vphi_k)&=&-2m\mu_{k+1}\ \sinh(\rho_k-\vphi_k)
\label{backlund3}\\
\d_{z_-}(\rho_k
-\vphi_k)&=&-\frac{2m}{\mu_{k+1}}\ \sinh(\rho_k+\vphi_k)
\label{backlund4}
\fin
Comparing with eqs.(\ref{EVi}) shows that the fields $\vphi_k$
and $\rho_k$ are connected through B\"acklund transformations as
follows~:
\def\tto#1{\  \stackrel{ B_{#1} }{\longrightarrow} \  }
\begin{eqnarray}
0 \tto{\mu_1} \vphi_1 \tto{-\mu_2} \rho_1 \tto{\mu_2} \vphi_2
\tto{-\mu_3}\cdots
\cdots\tto{\mu_{k-1}} \vphi_{k-1} \tto{-\mu_k} \rho_{k-1} \tto{\mu_k}
\vphi_k
\cdots\label{EVvii}
\fin
These B\"acklund transformations are for the fields satisfying the
sinh-Gordon model. But, since the auxiliary fields $\zeta$ in the
affine sinh-Gordon model, eqs.(\ref{Mo1},\ref{Mo2},\ref{Mo3}), are
determined from the field
$\vphi$,
the B\"acklund transformations (\ref{EVvii}) can be lifted to the
affine model.
Notice that in order to find the N-soliton $\tau$-function, we have
to apply $2N$ B\"acklund transformations, i.e. we apply sucessively
$B_{-\mu_k}$ and $B_{\mu_k}$ N times. This is in contrast with the
standard
 computation.

We now use the relation (\ref{EVvii}) to solve for the N-soliton
$\tau$-functions.

\proclaim Proposition.
The N-soliton $\tau$-functions $\tau^{(N)}_\pm$ as defined in
eq.(\ref{ETN})
are polynomials in the variables $X_k$, $k=1,\cdots, N$, with~:
\begin{eqnarray}
X_k = a_k \exp( 2m(\mu_kz_++ \mu_k^{-1} z_-))
\fin
They satisfy the following recursion relation~:
\begin{eqnarray}
\tau_\pm^{(N)}(X_k) = \tau_\pm^{(N-1)}(X_k)\ \pm\
X_{N}\ \tau_\pm^{(N-1)}(\beta^2_{Nk}X_k) \label{EVix}
\fin
This determines them uniquely. The solution of this recursion relations
are the $\tau$-functions (\ref{tau+-dev}).

We prove it in the Appendix D using the recursive B\"acklund
transformations
(\ref{EVvii}). As an intermediate step of the proof, we show that
the fields $\rho_k$ correspond to the ${k}$-soliton solutions
$\vphi_k$
but with a rescaling of the variables $X_j$. Namely~:
\begin{eqnarray}
\rho_k(X_j) = \vphi_{k}(\beta_{k+1;j}X_j)
\fin

Finally, notice that it is rather intriguing that the dressing
transformations which, by essence, are non-local and Lie-Poisson
actions are mapped into a set of B\"acklund transformations which
are more naturally connected to the local conserved currents.

\section{Relation with the Vertex operators.}

Another well known construction of the solitons solutions was given by
the
Kyoto school. It is ultimately expressed in terms of vertex operators.
We
show in the following sections that the previous analysis directly
leads to
the Japanese formulae \cite{DaJiKaMi81} when we use the level one
vertex  operator
representation of the affine $sl_2$ algebra.

\subsection{Vertex operator representations of $\hat{sl}_2$.}

We begin by a description of the level one representations of
$\hat{sl}_2$.
Following \cite{LeWi78}, we introduce oscillators $p_n$ for $n$
odd,
such that
\begin{eqnarray}
\relax [ p_m,p_n]=m \delta_{n+m,0} \nonumber
\end{eqnarray}
Assume $p_n^+ =p_{-n}$. The vacuum is defined by
\begin{eqnarray}
p_n |0> =0 ~~~~~~n>0 \nonumber
\end{eqnarray}
The associated normal ordering is defined by putting $p_n$, $n>0$ to
the right. Let
\begin{eqnarray}
Z(\lambda) = -i\sqrt{2} \sum_{n~odd} p_{-n} {\lambda^n \over n}
\nonumber
\end{eqnarray}
We have
\begin{eqnarray}
< Z(\lambda ) Z(\mu ) > =\log\( {\lambda + \mu \over \lambda - \mu }\)
\qquad ~|\lambda | > |\mu | \nonumber
\end{eqnarray}
The vertex  is defined by
\begin{eqnarray}
V(r,\lambda ) ={\textstyle{1\over 2}} \norm e^{ir Z(\lambda)} \norm
\nonumber
\end{eqnarray}
We have
\begin{eqnarray}
V(r,\lambda ) V(s,\mu )
= \left( {\lambda -\mu \over \lambda + \mu } \right)^{rs}  \norm
V(r,\lambda) V(s,\mu)
 \norm ~~~
|\lambda |> |\mu | \nonumber
\end{eqnarray}
The level one vertex operator representations of the Lie algebra
$\hat{sl}_2$
are obtained as follows:
\begin{eqnarray}
\sum_{n~odd} \lambda^{-n} (E_+^n + E_-^n) &=& {i\over \sqrt 2} \lambda
{d
\over
d\lambda } Z(\lambda ) \nonumber\\
\sum_{n~even} \lambda^{-n} H^n + \sum_{n~odd} \lambda^{-n} (E_+^n -
E_-^n )&=& \pm\ V(\lambda)
\label{verop}
\end{eqnarray}
where $V(\lambda)$ denotes the vertex
 \begin{eqnarray}
V(\lambda )= V(-\sqrt{2},\lambda)= {\textstyle{1\over 2}}\norm e^{- i
\sqrt{2} Z(\lambda ) } \norm
= {\textstyle{1\over 2}} \norm  e^{ i \sqrt{2} Z(- \lambda ) }\norm
\nonumber
\end{eqnarray}
The representation with highest weight $\La^+$
($\La^-$) corresponds to the plus (minus) sign in eq.(\ref{verop}).
It is then a standard calculation to verify that we have obtained
a representation of the affine $sl_2$ algebra at level one in the
principal
gradation. In particular we have
\begin{eqnarray}
\relax [ E_+^n + E_-^n  , V(\lambda) ] =- 2 \lambda^n
V(\lambda)  \label{root}
\end{eqnarray}
Finally, remark that $V(\la)^2=0$ inside any expectation value.

\subsection{$\tau$-functions and Vertex operators.}

We have shown in the previous sections that the element $g_-^{-1} g_+$
has
very special commutation properties with $e^{-mz_+ {\cal E}_+}$ and
$e^{-mz_- {\cal E}_-}$. In fact, commuting these factors simply
reintroduce
the $z_\pm$ evolution of the solitons.
\begin{eqnarray}
a_i \longrightarrow X_i = a_i e^{2m (\mu_i z_+ + \mu_i^{-1} z_- )}
\nonumber
\end{eqnarray}
In view of eq.(\ref{root}) we see that the vertex operator has the same
property
\begin{eqnarray}
e^{-mz_- {\cal E}_-}e^{-mz_+ {\cal E}_+} V(\mu_i) e^{mz_+ {\cal E}_+}
e^{mz_- {\cal E}_-} =  e^{2m (\mu_i z_+ + \mu_i^{-1} z_- )} V(\mu_i)
\nonumber
\end{eqnarray}
So, we expect that $g= g_-^{-1} g_+$ has a simple expression in terms
of
$V$. This is indeed the case.
\proclaim Proposition.
In the level one representations, we have~:
\begin{eqnarray}
g = (1+2 a_1 V(\mu_1))(1+2a_2 V(\mu_2))\cdots (1+2 a_N V(\mu_N))
\label{japon}
\end{eqnarray}
where $g=g_-^{-1}g_+$ is defined in eq.(\ref{EIVi}-\ref{g-k}).
As a consequence, we get the following alternative  expression for the
$\tau$-functions
\begin{eqnarray}
\frac{\tau_\pm^{(N)}}{\tau_0}(z_\pm^{(s)}) = <\Lambda^\pm \vert
\prod_{i=1}^N (1+ 2 X_i V(\mu_i)) \vert
\Lambda^\pm > \non
\end{eqnarray}
where
\begin{eqnarray}
X_i = a_i e^{2m \left( \sum_s \mu_i^s z_+^{(s)} + \mu_i^{-s} z_-^{(s)}
\right)
}
\nonumber
\end{eqnarray}

\proof
To prove these formulae, we remark that the meaning of the elements
$V_\mu^{(\pm)}$ discovered in our study of the
solitons in the sinh-Gordon model is now clear. They are just the
projections of the vertex operator $V(\mu)$ on ${\cal B}_\pm$
respectively.
\begin{eqnarray}
V_\mu^{(-)} + V_\mu^{(+)} = V(\mu) \nonumber
\end{eqnarray}
moreover the projections of $V_\mu^{(\pm)}$ on ${\cal H}$ are equal.
In other words, $V^{(\pm)}_\mu$ are the solutions of the factorization
of
$V(\mu)$ corresponding to the dressing problem. Therefore
\begin{eqnarray}
V_\mu^{(-)} &=&~~ {1\over 2} \oint_{|\lambda |>|\mu |} {d\lambda \over
2i\pi
\lambda } {\lambda + \mu \over \lambda - \mu } V(\lambda ) \nonumber \\
V_\mu^{(+)} &=&- {1\over 2} \oint_{|\lambda |<|\mu |} {d\lambda \over
2i\pi
\lambda } {\lambda + \mu \over \lambda - \mu } V(\lambda ) \nonumber
\end{eqnarray}
These operators satisfy the remarkable identity~:
\begin{eqnarray}
e^{f V_\mu^{(-)}} e^{f V_\mu^{(+)}}=  \cosh {f\over 2}
+2 \sinh {f\over 2} ~V(\mu)  \label{B1}
\end{eqnarray}
We prove it in the Appendix E. It relies on the relation:
\begin{eqnarray}
V^{(+)}_\mu\,^2 +2 V^{(-)}_\mu V^{(+)}_\mu
+ V^{(-)}_\mu\,^2 = \inv{4}\non
\end{eqnarray}
which is satisfied in the level one vertex operator representations.

 Our claim in the one-soliton case immediately follows from this
 formula.
Remember that for the one-soliton case we had~:
\begin{eqnarray}
g_-^{-1}g_+ = e^{\half\log(1-a^2)K} e^{f V_\mu^{(-)}} e^{f V_\mu^{(+)}}
\non
\end{eqnarray}
with $f=\log(1+a)/(1-a)$. Therefore, since $K=1$,
\begin{eqnarray}
g_-^{-1}g_+ = 1 + 2a V(\mu)
\label{EJap1}
\end{eqnarray}

In the general N-soliton case, we have to evaluate,
\begin{eqnarray}
g_-^{-1}g_+ = g_-^{-1}(1)  g_-^{-1}(2) \cdots  g_-^{-1}(N)\cdot
g_+(N) g_+(N-1) \cdots g_+(1) \nonumber
\end{eqnarray}
where all the factors are defined in eq.(\ref{g-k},\ref{g+k}).
The middle terms can be computed from the identity (\ref{B1}), we
have~:
\begin{eqnarray}
g_-^{-1}(N)\cdot g_+(N) = \vev{E(N)| W(N)} e^{g_N(0)K} \non
\end{eqnarray}
where we have introduced a convenient bra-ket notation by denoting
$\bra{E(k)}$ the line vector with operator entries
\begin{eqnarray}
\bra{E(k)}= \left( 1,V(\mu_k) , V(-\mu_k ), {i\over \sqrt{2}} \mu_k
{d\over
d \mu_k } Z(\mu_k ) \right) \nonumber
\end{eqnarray}
and $\ket{W(k)}$ the column vector
\begin{eqnarray}
\ket{W(k)} = \pmatrix{\cosh {f_k \over 2 } \cr 2 \sinh {f_k \over 2 }
\cr 0
\cr 0}
\nonumber
\end{eqnarray}
Next one has to commute $g_+(N-1)$ through $\bra{E(N)}$.
In the Appendix F, we prove the relations~:
\begin{eqnarray}
\bra{E(k)}\ g_+(j) = g_+(j)\ \bra{E(k)}\ R(j,k)\label{Eg+}
\end{eqnarray}
with $R(j,k)$ a $C$-number matrix explicitly computed in the Appendix
F.
As a result, the commutation of $g_+(N-1)$ gives the factors~:
\begin{eqnarray}
\vev{E(N-1)|W(N-1)}\ \bra{E(N)}R(N-1,N)\ket{W(N)}
e^{(g_{N-1}(0)+g_N(0))K} \non
\end{eqnarray}
The general case is now clear, we obtain~:
\begin{eqnarray}
g=g_-^{-1} g_+ = g(1) g(2) \cdots g(N) \nonumber
\end{eqnarray}
where
\begin{eqnarray}
g(k) = \bra{E(k)} \left( \prod_{j=1}^{k-1} R(j,k) \right) \ket{W(k)}
e^{g_K(0)K}
\nonumber
\end{eqnarray}
The next step consists in showing that the $f_k$ and $h_k$ are such
that:
\begin{eqnarray}
R(1,k) R(2,k) \cdots R(k-1 ,k ) \ket{W(k)} =\Delta_k \pmatrix{1 \cr 2
X_k
\cr 0 \cr 0 }
\label{nso01}
\end{eqnarray}
where $\Delta_k$ are some coefficients and $X_k$ is the variable
eq.(\ref{Xi}) of the $k^{th}$-soliton.
We relegate the proof of this assertion in the Appendix G .
As a consequence, we get for $z_+ =z_-=0$~:
\begin{eqnarray}
g = \({\prod_{k=1}^N\Delta_k e^{g_k(0)K}}\)
(1+2 a_1 V(\mu_1))(1+2a_2 V(\mu_2))\cdots (1+2 a_N V(\mu_N)) \nonumber
\end{eqnarray}
We also find the consistency relation $\prod_k\Delta_ke^{g_k(0)}=1$
which can
 be checked directly.
It is now trivial to recalculate the $\tau$-functions from this
expression.
We of course get back eq.(\ref{tau+-dev}).\cqfd

\section{Appendix A: Determination of $g_\pm$ in the affine group.}

Here we determine the gauge transformation $g_\pm(x,t)$ in the affine
group for
the one-soliton case. In the course of this computation we will
find the first few constraints characterizing the vacuum orbit.
Let us consider $g_- (x,t)$. It is determined by the equations
(\ref{gauge+}) and (\ref{gauge-}).
We look for $g_-(x,t)$ as a graded product
\begin{eqnarray}
g_{-}(x,t)=g_0 g_{-1} g_{-2} \cdots \non
\end{eqnarray}
where the lower indices refer
to the degree in the affine group in the principal gradation.
Comparing the terms of degree one in eq.(\ref{gauge+}) and of degree
zero in eq.(\ref{gauge-}), we find
$g_0=e^{\Phi_{sol}-\Phi_{vac}}$ in agreement with the definition of
the factorization problem.
Next, we set $g_{-1}=e^{X_{-1}}$, with $degree(X_{-1})=-1$.
Comparing the terms of degree zero in eq.(\ref{gauge+}) we obtain
the condition for $X_{-1}$~:
\begin{eqnarray}
m[X_{-1},{\cal E}_+]=2\partial_{z_+} (\Phi_{sol} -\Phi_{vac} )
 \nonumber
\end{eqnarray}
Its solution is
$X_{-1}= X_{-\alpha_1} E_{-\alpha_1}+ X_{-\alpha_2} E_{-\alpha_2}$
 with~:
\begin{eqnarray}
X_{-\alpha_1}&=&-{1\over{2m}}\partial_{z_+}(\varphi_{sol} +\zeta_{sol}
-\zeta_{vac} ) \nonumber \\
X_{-\alpha_2}&=& -{1\over{2m}}\partial_{z_+}(-\varphi_{sol}
+\zeta_{sol} -\zeta_{vac} ) \nonumber
\end{eqnarray}
The next level is more interesting. We set
$g_{-2}=e^{X_{-2}}$, with $degree(X_{-2})=-2$.
The equation determining $X_{-2}$ is
\begin{eqnarray}
m[X_{-2},{\cal E}_+] &=& \partial_{z_+}X_{-1}-
[X_{-1}, \partial_{z_+}\Phi_{sol} ]
\nonumber
\end{eqnarray}
There is only one root at level $-2$, therefore,
$X_{-2}= X_{-\alpha_1 -\alpha_2} [E_{-\alpha_1},E_{-\alpha_2}]$.
The equation for $X_{-\alpha_1 -\alpha_2}$ reads
\begin{eqnarray}
2m X_{-\alpha_1 -\alpha_2}(E_{-\alpha_1}-E_{-\alpha_2})&=&
(\partial_{z_+} X_{-\alpha_1} - \partial_{z_+} \varphi_{sol}
X_{-\alpha_1})E_{-\alpha_1}
\nonumber \\
&&+(\partial_{z_+} X_{-\alpha_2} + \partial_{z_+} \varphi_{sol}
    X_{-\alpha_2}) E_{-\alpha_2}
\nonumber
\end{eqnarray}
These are two equations for one unknown. Therefore, we get the
constraint
\begin{eqnarray}
(\partial_{z_+} \varphi_{sol})^2 - \partial_{z_+}^2 (\zeta_{sol} -
\zeta_{vac} ) =0
\nonumber
\end{eqnarray}
This constraint can easily be checked for one and two solitons. This is
the first
of an infinite set of constraints (which will appear at every even
degree), characterizing the orbit of the vacuum. To proceed along this
line would rapidly becomes
untractable. Fortunately, at this point the computation can be done in
the loop
group. Indeed, we notice that all central terms occur with
$g_{-1}$ and they have been taken care of. Defining
\begin{eqnarray}
g_- (x,t) = e^{{1\over 4} (\zeta_{sol} -\zeta_{vac})K}~\widetilde{g}_-
(x,t) \nonumber
\end{eqnarray}
one can simply
forget the central extension and work in the loop group.
The equation for $\widetilde{g}_-$ in the loop group reads
\begin{eqnarray}
\widetilde{g}_-(x,t)\left[ \partial_{z_+} +m {\cal E}_+ \right]
\widetilde{g}_-^{-1}(x,t) =
\textstyle{1\over 2} \partial_{z_+}\varphi_{sol} H +
m e^{\half \varphi_{sol} ad~H} {\cal E}_+ \nonumber
\end{eqnarray}
or more explicitly eq.(\ref{g-tilde}).
At this point, we make contact with the computation in the loop group
as explained in section(5.1).

\section{Appendix B. The elements $V_\mu^{(+)}$ and $V_\mu^{(-)}$. }

We list some formulae concerning the commutation relations between
$V_\mu^{(\pm)}$ and ${\cal E}_\pm$. Let
\begin{eqnarray}
U_\mu^{(\pm)}={1\over 2} \left( 1-{\mu^{\mp 1} \over 2 } ad~ {\cal
E}_\pm \right) H
\nonumber
\end{eqnarray}
then
\begin{eqnarray}
ad~ V_\mu^{(\pm)}\cdot {\cal E}_\pm &=& 2\mu^{\pm 1} \left(
V_\mu^{(\pm)}
- U_\mu^{(\pm )} \right) \nonumber\\
ad~ V_\mu^{(\pm)}\cdot {\cal E}_\mp &=& 2\mu^{\mp 1}  \left(
V_\mu^{(\pm)}
+ U_\mu^{(\mp )} \right)
\nonumber
\end{eqnarray}
Therefore, defining $V_\mu =  V_\mu^{(+)}+ V_\mu^{(-)}$,
we see that
$ad~ V_\mu \cdot {\cal E}_\pm = 2 \mu^{\pm 1} V_\mu$. See also
\cite{OTxx}.
Strictly speaking, in the loop representation the element $V_\mu$ is
not defined,
but it is well defined for instance in the level one representation.
Next, one
has
\begin{eqnarray}
ad~  V_\mu^{(\pm)} \cdot U_\mu^{(\pm )}
&=& -{\mu^{\mp 1}\over 2} {\cal E}_\pm \nonumber\\
ad~  V_\mu^{(\pm)} \cdot U_\mu^{( \mp )}
&=& { \mu^{\pm 1}\over 2} {\cal E}_\mp  \mp  \frac{K}{2} \nonumber
\end{eqnarray}
 From this we get
\begin{eqnarray}
e^{-f ad~V_\mu^{(\pm)}}{\cal E}_\mp &=& \cosh(f) {\cal E}_\mp
-2\mu^{\mp 1} \sinh(f) \left(
V_\mu^{(\pm)} +U_\mu^{(\mp )} \right)
\mp \mu^{\mp 1} [\cosh(f)-1]K \nonumber \\
e^{-f ad~ V_\mu^{(\pm)}}  {\cal E}_\pm &=& \cosh(f) {\cal E}_\pm -
2\mu^{\pm 1} \sinh(f)  \left(
V_\mu^{(\pm)} - U_\mu^{(\pm)} \right)
\label{three}
\end{eqnarray}
also
\begin{eqnarray}
e^{-f ad~V_\mu^{(\pm)}}U_\mu^{(\mp )} &=&
-{\mu^{\pm 1}\over 2} \sinh(f) {\cal E}_\mp + (\cosh(f)
-1)V_\mu^{(\pm)}
+ \cosh(f) U_\mu^{(\mp)}
 \pm  \sinh(f){K\over 2}
\nonumber \\
e^{-f ad~V_\mu^{(\pm)}}U_\mu^{\pm} &=&
{\mu^{\mp 1}\over 2} \sinh(f) {\cal E}_\pm -( \cosh(f) -1)V_\mu^{(\pm)}
+ \cosh(f) U_\mu^{(\pm)}
\label{fivep}
\end{eqnarray}

\section{Appendix C: Solution of the N-soliton dressing problem.}

We look for elements $Y_+(k,z_+)$ such that
\begin{eqnarray}
Y_+(k,z_+) e^{f_k(0) V_{\mu_k}^{(-)}} e^{{1\over 2}h_k(0) H
}e^{{1\over
2}g_k(0)K}
=
e^{f_k(z_+) V_{\mu_k}^{(-)}}
e^{{1\over 2}h_k(z_+) H }e^{{1\over 2}g_k(z_+)K} Y_+(k+1,z_+)
\nonumber
\end{eqnarray}
or equivalently
\begin{eqnarray}
\partial_{z_+} Y_+(k+1,z_+) Y_+(k+1,z_+)^{-1}
& =& \label{four} \\
e^{-{1\over 2} h_k ad~H }
\Big\{
e^{-f_k ad~ V_{\mu_k}^{(-)}} [ \partial_{z_+} Y_+(k,z_+)
Y_+(k,z_+)^{-1} ]
&-&\partial_{z_+} f_k~ V_{\mu_k}^{(-)} -{1\over 2}\partial_{z_+} h_k H
-{1\over 2}\partial_{z_+} g_k K \Big\}
\nonumber
\end{eqnarray}
We now make the crucial remark that the above equations can be
satisfied if we
take the $Y_+(k,z_+)$'s such that (compare with eq.(\ref{elem2}) )
\begin{eqnarray}
\partial_{z_+} Y_+(k,z_+) Y_+(k,z_+)^{-1} = -m\cosh\phi_k~ {\cal E}_+
+2m \mu_k \sinh\phi_k~ U_{\mu_k}^{(+)}\equiv {\cal Y}_+(\phi_k, \mu_k)
\label{five}
\end{eqnarray}
The following commutation relations hold
\begin{eqnarray}
e^{-f_k ad~ V_{\mu_k}^{(-)}}~  {\cal Y}_+(\phi_k, \mu_k)&=&
{\cal Y}_+(\phi_k + f_k , \mu_k)
+2m\mu_k \left[ \sinh(\phi_k +f_k)-\sinh (\phi_k ) \right]
V_{\mu_k}^{(-)} \nonumber \\
&&- m\mu_k  \left[ \cosh (\phi_k + f_k) - \cosh (\phi_k) \right] K
\nonumber
\end{eqnarray}
\begin{eqnarray}
e^{-{1\over 2}h_k ad~ H}~  {\cal Y}_+(\phi_k +f_k , \mu_k)&=&
{\cal Y}_+(\phi_k + f_k+h_k , \mu_{k+1})\non \\
&& + \left[ m\mu_k \sinh (\phi_k + f_k) -
m\mu_{k+1} \sinh (\phi_k + f_k +h_k) \right] H  \nonumber
\end{eqnarray}
Using the first of these relations, eq.(\ref{four}) becomes
\begin{eqnarray}
\partial_{z_+} Y_+(k+1,z_+) Y_+(k+1,z_+)^{-1} &=&
e^{-{1\over 2} h_k ad~H} \Big\{ \CY_+(\phi_k+f_k,\mu_k)
 - {1\over 2} \partial_{z_+} h_k H      \non\\
&-& \Big[ \partial_{z_+} f_k~ + 2m\mu_k \sinh~(\phi_k)
 -2m \mu_k \sinh~(\phi_k +f_k) \Big]  V_{\mu_k}^{(-)} \non\\
&-&{1\over 2}\Big(\partial_{z_+} g_k + 2m \mu_k [ \cosh (\phi_k + f_k)
-\cosh (\phi_k) ]
\Big) K \Big\}
\nonumber
\end{eqnarray}
Projecting this equation on ${\cal G}_-$ and on the central element, we
get
\begin{eqnarray}
\partial_{z_+} f_k &=& 2m\mu_k [ \sinh(\phi_k + f_k )- \sinh(\phi_k)]
\label{une}\\
\partial_{z_+} g_k &=& -2m\mu_k  [ \cosh (\phi_k + f_k) -\cosh\phi_k ]
\nonumber
\end{eqnarray}
while the projection on ${\cal G}_+$ gives
\begin{eqnarray}
\partial_{z_+} Y_+(k+1,z_+) Y_+(k+1,z_+)^{-1} = e^{-{1\over 2} h_k ad~
H} \Big\{
\CY_+(\phi_k+ f_k,\mu_k)
- {1\over 2} \partial_{z_+} h_k H \Big\}
\label{six}
\end{eqnarray}
Next, using the commutation relations between $H$ and $\CY_+$,
eq.(\ref{six}) becomes
\begin{eqnarray}
\partial_{z_+} Y_+(k+1) Y_+(k+1)^{-1} &=&
{\cal Y}_+(\phi_k + f_k+h_k , \mu_{k+1}) \nonumber \\
&+& \left[m \mu_k \sinh (\phi_k + f_k) -
m\mu_{k+1} \sinh (\phi_k + f_k +h_k)- \half \d_{z_+} h_k \right] H
\non
\end{eqnarray}
Choosing the $z_+$ evolution of $h_k$ as
\begin{eqnarray}
\partial_{z_+} h_k=2m \mu_k \sinh (\phi_k + f_k) -2m\mu_{k+1} \sinh
(\phi_k + f_k +h_k) \label{trois}
\end{eqnarray}
we recover an expression of the form eq.(\ref{five}) with
\begin{eqnarray}
\phi_{k+1} &=& \phi_k + f_k +h_k \label{deux}
\end{eqnarray}
Eq.(\ref{une},\ref{deux},\ref{trois}) describe the solution of the
dressing
problem.

Next, we relate these equations to the B\"acklund transformations. Let
us set:
\begin{eqnarray}
f_k &=& \varphi_{k-1} -\varphi_k ~~~~~~~~~~~\varphi_0 =0 \nonumber \\
h_k &=& \rho_k - \rho_{k-1} ~~~~~~~~~~~~\rho_0 =0 \nonumber
\end{eqnarray}
Eq.(\ref{deux}) gives
$\phi_k = \rho_{k-1} - \varphi_{k-1}$
and therefore
$\phi_k + f_k = \rho_{k-1} - \varphi_k$ .
The recursion equations (\ref{une},\ref{trois}) become
\begin{eqnarray}
\partial_{z_+} (\varphi_{k-1} -\varphi_k ) &=& 2m \mu_k [
\sinh(\rho_{k-1}
-\varphi_k ) -
\sinh~(\rho_{k-1}-\varphi_{k-1})] \nonumber \\
\partial_{z_+} ( \rho_k -\rho_{k-1} ) &=& 2m \mu_k \sinh (\rho_{k-1}
-\varphi_k )
-2m \mu_{k+1} \sinh(\rho_k - \varphi_k ) \nonumber
\end{eqnarray}
from which we deduce
\begin{eqnarray}
\partial_{z_+} (\varphi_{k-1} + \rho_{k-1})+ 2m\mu_k \sinh(\rho_{k-1}
-\varphi_{k-1}
)&=&
\partial_{z_+} (\varphi_{k} + \rho_{k})+ 2m\mu_{k+1} \sinh (\rho_{k}
-\varphi_{k} )
\nonumber
\\
\partial_{z_+} (\varphi_{k} + \rho_{k-1})+ 2m\mu_k \sinh(\rho_{k-1}
-\varphi_{k}
)&=&
\partial_{z_+} (\varphi_{k+1} + \rho_{k})+ 2m\mu_{k+1} \sinh(\rho_{k}
-\varphi_{k+1}
)
\nonumber
\end{eqnarray}
or
\begin{eqnarray}
\partial_{z_+} (\varphi_{k} + \rho_{k})&=& 2m\mu_{k+1} \sinh(\varphi_k
-\rho_{k} )
\label{quatre}
\\
\partial_{z_+} (\varphi_{k} + \rho_{k-1})&=&  2m\mu_k
\sinh(\varphi_{k}-\rho_{k-1})
\label{cinq}
\end{eqnarray}
Similarly, one shows that the factorization problem for $g_+$ is
solved in the same way. The $z_-$ dependence can also be treated
in a similar way.

\section{Appendix D: Determination of the $\tau$-functions. }

In order to solve the sucessive set of B\"acklund transformations
eqs.(\ref{backlund1},
\ref{backlund2},\ref{backlund3},\ref{backlund4})
by using the `tangent rule', we have to introduce a larger set of
transformations. Before plunging into the general case, let us first
consider the two-soliton case. Starting from the zero-soliton solution
$\Phi_0$ and from two one-soliton solutions, $\Phi_1$ and $\Phi_2$,
with parameter $a_i,\ \mu_i$, $i=1,\ 2$,
we consider the following B\"acklund transformations~:
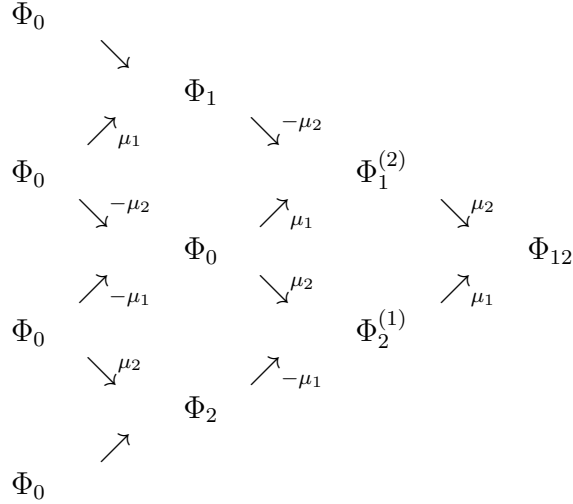
\begin{eqnarray}
\matrix{
\Phi_0 & & & & & & \cr
 & \searrow& & & & & \cr
 & & \Phi_1 & & & & \cr
 & \nearrow_{\mu_1} & & \searrow^{-\mu_2} & & &\cr
 \Phi_0 & & & &\Phi_1^{(2)} & & \cr
 & \searrow^{-\mu_2} & & \nearrow_{\mu_1} & &\searrow^{\mu_2}& \cr
 & & \Phi_0 & & & & \Phi_{12}\cr
 & \nearrow_{-\mu_1} & & \searrow^{\mu_2} & &\nearrow_{\mu_1}& \cr
 \Phi_0 & & & & \Phi_2^{(1)} & & \cr
 & \searrow^{\mu_2} & & \nearrow_{-\mu_1} & & &\cr
 & & \Phi_2 & & & & \cr
 & \nearrow& & & & & \cr
\Phi_0 & & & & & & \cr}\non
\end{eqnarray}
The upper line is formed with  the B\"acklund transformations involved
in the
dressing problem, eq.(\ref{EVvii}). The lower line is the same but with
the exchange of the indices 1 and 2. Using the `tangent rule' applied
to the $\tau$-functions (\ref{EVv},\ref{EVv'}), we first compute
$\Phi_1^{(2)}$ from
the knowledge of $\Phi_0$ and $\Phi_1$. By exchange of $\mu_1$ and
$\mu_2$, it also gives $\Phi_2^{(1)}$. Then we compute $\Phi_{12}$
from $\Phi_0$, $\Phi_1^{(2)}$ and $\Phi_2^{(1)}$. We find~:
\begin{eqnarray}
\tau_\pm(\Phi_1^{(2)})&=&1\pm \beta_{21} X_1 \non\\
\tau_\pm(\Phi_{12})&=&1\pm  X_1 \pm X_2 + \beta_{12}^2 X_1 X_2 \non
\end{eqnarray}
as it should be.

We now proceed by proving by induction that :
\begin{eqnarray}
\tau(\Phi_{1\cdots N}) &=& \tau_{1\cdots N}(X_j) \non\\
\tau(\Phi_{1\cdots N}^{(a)}) &=& \tau_{1\cdots N}(\beta_{a;j}X_j)
\non\\
\tau(\Phi_{1\cdots N}^{(a;b)}) &=& \tau_{1\cdots
N}(\beta_{b;j}\beta_{a;j}X_j) \label{AAi}
\end{eqnarray}
The lower indices on $\Phi$ refer to the number of soliton parameters
$X_j$,
and the upper indices refer to the shift of the factors $X_j$ in the
$\tau$-functions.
We assume that the fields are connected via B\"acklund transformations
as follows~:
\begin{eqnarray}
\matrix{
\cdots &\Phi_{1\cdots N-1} & & & & & & & & \cr
 & & \searrow^{-\mu_N}& & & & & & & \cr
 & & & \Phi_{1\cdots N-1}^{(N)}& & & & & & \cr
 & & \nearrow_{\mu_1} & & \searrow^{\mu_N} & & & & &\cr
 \cdots &\Phi_{2\cdots N-1}^{(1;N)} & & & &\Phi_{1\cdots N} & & & & \cr
 & & \searrow^{\mu_N} & & \nearrow_{\mu_1} & &\searrow^{-\mu_{N+1}}& &
 &\cr
 & & & \Phi_{2\cdots N}^{(1)} & & & & \Phi_{1\cdots N}^{(N+1)} & & \cr
 & & \nearrow_{-\mu_1} & & \searrow^{-\mu_{N+1}} & &\nearrow_{\mu_1}&
	&\searrow^{\mu_{N+1}} &\cr
\cdots&\Phi_{2\cdots N} & & & & \Phi_{2\cdots N}^{(1;N+1)} & & &
&\Phi_{1\cdots N+1} \cr
&  & \searrow^{-\mu_{N+1}} & & \nearrow_{-\mu_1} &
&\searrow^{\mu_{N+1}}&
	&\nearrow_{\mu_1} &\cr
 & & & \Phi_{2\cdots N}^{(N+1)} & & & & \Phi_{2\cdots N+1}^{(1)} & &
 \cr
 & & \nearrow_{\mu_2} & & \searrow^{\mu_{N+1}} & &\nearrow_{-\mu_1}& &
 &\cr
 \cdots&\Phi_{2\cdots N-1}^{(2;N+1)} & & & &\Phi_{2\cdots N+1} & & & &
 \cr
 & & \searrow^{\mu_{N+1}} & & \nearrow_{\mu_2} & & & & &\cr
 & & & \Phi_{3\cdots N+1}^{(2)}& & & & & & \cr
 & & \nearrow_{-\mu_2}& & & & & & & \cr
\cdots & \Phi_{3\cdots N+1} & & & & & & & & \cr}
  \label{AAii}
\end{eqnarray}

As the recursion hypothesis we assume that  the relations (\ref{AAi})
have been
proved for all the fields in (\ref{AAii}) but the last four ones:
$\Phi_{2\cdots N}^{(1;N+1)}$, $\Phi_{1\cdots N}^{(N+1)}$,
$\Phi_{2\cdots N+1}^{(1)}$ and $\Phi_{1\cdots N+1}$.
Then, we have to prove that the relations (\ref{AAi}) hold also for
these
four fields.
The field $\Phi^{(1;N+1)}_{2\cdots N}$ is computed from $\Phi_{2\cdots
N}$,
$\Phi_{2\cdots N}^{(1)}$ and $\Phi_{2\cdots N}^{(N+1)}$ using the
`tangent rule';
Then $\Phi_{1\cdots N}^{(N+1)}$ is computed from $\Phi_{2\cdots
N}^{(1)}$,
$\Phi_{1\cdots N}$ and $\Phi_{2\cdots N}^{(1;N+1)}$, and similarly for
$\Phi_{2\cdots N+1}^{(1)}$; Finally, $\Phi_{1\cdots N+1}$ is computed
from
$\Phi_{2\cdots N}^{(1;N+1)}$, $\Phi_{1\cdots N}^{(N+1)}$,  and
$\Phi_{2\cdots N+1}^{(1)}$. In all these sucessive computations,
proving
the relations (\ref{AAi}) reduces in proving the two following
identities~:
\begin{eqnarray}
&&\tau_+^{(N)}(\beta_{a;k}\beta_{b;k} X_k) \tau_-^{(N)}(X_k) +
\tau_-^{(N)}(\beta_{a;k}\beta_{b;k} X_k) \tau_+^{(N)}(X_k) \non\\
&&= \tau_+^{(N)}(\beta_{a;k} X_k) \tau_-^{(N)}(\beta_{b;k}X_k) +
\tau_-^{(N)}(\beta_{a;k} X_k) \tau_+^{(N)}(\beta_{b;k}X_k) \label{EBi}
\end{eqnarray}
\begin{eqnarray}
\beta_{ab}\Big[\tau_+^{(N)}(\beta_{a;k}\beta_{b;k} X_k)
\tau_-^{(N)}(X_k) &-&
\tau_-^{(N)}(\beta_{a;k}\beta_{b;k} X_k) \tau_+^{(N)}(X_k) \Big]\non\\
= \tau_+^{(N)}(\beta_{a;k} X_k) \tau_-^{(N)}(\beta_{b;k}X_k) &-&
\tau_-^{(N)}(\beta_{a;k} X_k) \tau_+^{(N)}(\beta_{b;k}X_k) \label{EBii}
\end{eqnarray}
We demonstrate them also by induction assuming that the
$\tau$-functions
satisfy the following recursion relation~:
\begin{eqnarray}
\tau_\pm^{(N)}(X_k) = \tau^{(N-1)}_\pm(X_k) \pm X_N
\tau_\pm^{(N-1)}(\beta^2_{N;k}X_k)\non
\end{eqnarray}
First let us start with (\ref{EBi}). We consider the left hand side
as a function of $\mu_a$, denoted $F(\mu_a)$.
It has simple poles for $\mu_a=-\mu_k$ and is finite at
$\mu_a\to\infty$.
Therefore, as a function of $\mu_a$
\begin{eqnarray}
F(\mu_a) = \sum_{k=1}^N\ \frac{Res_k}{\mu_a+\mu_k} +
F(\mu_a=\infty) \non
\end{eqnarray}
 Since $\beta_{a;k}\to 1$ for $\mu_a\to\infty$, the relation
 (\ref{EBi})
 is
 obviously satisfied at $\mu_a=\infty$. Hence, we only have to show
 that the
 residues at $\mu_a=-\mu_k$ of the left hand side and of the right hand
 side coincide. By symmetry, we can choose to only pick up the residue
 at
$\mu_a=-\mu_N$.Using the fact that~:
\begin{eqnarray}
\beta_{a;k}\big\vert_{\mu_a=-\mu_N} = \beta_{Nk}^{-1} \non
\end{eqnarray}
and the recursion formula for the $\tau$-functions, the equality of the
residue at $\mu_a=-\mu_N$ is equivalent to~:
\begin{eqnarray}
&& \beta_{b;N}\Big[\tau_+^{(N-1)}(\beta_{b;k}\beta_{N;k} X_k)
\tau_-^{(N)}(X_k) -
\tau_-^{(N-1)}(\beta_{b;k}\beta_{N;k} X_k) \tau_+^{(N)}(X_k)
\Big]\non\\
&=& \tau_-^{(N-1)}(\beta_{N;k} X_k) \tau_+^{(N)}(\beta_{b;k}X_k) -
\tau_+^{(N-1)}(\beta_{N;k} X_k) \tau_-^{(N)}(\beta_{b;k}X_k)
\label{Eiii}
\end{eqnarray}
Next, using once again the recursion formula for the $\tau$-functions,
we expand the relation (\ref{Eiii}) in $X_N$. It is linear in $X_N$.
The terms linear in $X_N$ are (after a rescaling $X_k\to
\beta^{-1}_{N;k}X_k$)~:
\begin{eqnarray}
&&\tau_+^{(N-1)}(\beta_{b;k}\beta_{N;k} X_k) \tau_-^{(N-1)}(X_k) +
\tau_-^{(N-1)}(\beta_{b;k}\beta_{N;k} X_k) \tau_+^{(N-1)}(X_k) \non\\
&=& \tau_+^{(N-1)}(\beta_{b;k} X_k) \tau_-^{(N-1)}(\beta_{N;k}X_k) +
\tau_-^{(N-1)}(\beta_{b;k} X_k) \tau_+^{(N-1)}(\beta_{N;k}X_k)
\label{EBip}
\end{eqnarray}
The terms independent in $X_N$ are~:
\begin{eqnarray}
\beta_{b;N}\Big[\tau_+^{(N-1)}(\beta_{b;k}\beta_{N;k} X_k)
\tau_-^{(N-1)}(X_k) &-&
\tau_-^{(N-1)}(\beta_{b;k}\beta_{N;k} X_k) \tau_+^{(N-1)}(X_k)
\Big]\non\\
= \tau_-^{(N-1)}(\beta_{b;k} X_k) \tau_+^{(N-1)}(\beta_{N;k}X_k) &-&
\tau_+^{(N-1)}(\beta_{b;k} X_k) \tau_-^{(N-1)}(\beta_{N;k}X_k)
\label{Eiip}
\end{eqnarray}
These two last relations are the recursion relations (\ref{EBi}) and
(\ref{EBii}) but one step before, i.e. for the $(N-1)$-soliton
$\tau$-functions. The relation (\ref{EBii}) is proved in the same way.
Hence, the recursion is proved~: the $\tau$-functions
which satisfy the recursion formula solve the B\"acklund
transformations
(\ref{AAii}).

\section{Appendix E. The one soliton vertex operator.}

We prove eq.(\ref{B1}). We have
\begin{eqnarray}
e^{f V_\mu^{(-)}} e^{f V_\mu^{(+)}} =\sum_{n=0}^\infty
{f^n \over n! } N(V_\mu^{(-)} + V_\mu^{(+)} )^n \nonumber
\end{eqnarray}
where the symbol $N()$ means writing $V_\mu^{(-)}$ on the left.
Let
\begin{eqnarray}
{\cal I}={1\over 4} {d\lambda_1 \over 2i\pi \lambda_1}~{d\lambda_2
\over 2i\pi
\lambda_2}~ {\lambda_1 +\mu \over \lambda_1 - \mu}~{\lambda_2 +\mu
\over
\lambda_2 - \mu}~ \left( {\lambda_1 -\lambda_2 \over \lambda_1 +
\lambda_2}\right)^2~ \norm V(\lambda_1 ) V(\lambda_2 ) \norm \nonumber
\end{eqnarray}
We have
\begin{eqnarray}
{V_\mu^{(-)}}^2 &=&~~~~ \oint \oint_{|\lambda_1|>|\lambda_2|>|\mu |}
{}~~{\cal I}
\nonumber \\
{V_\mu^{(+)}}^2 &=& ~~ ~~\oint \oint_{|\mu |>|\lambda_1|>|\lambda_2|}
{}~~{\cal I}
\nonumber \\
V_\mu^{(-)} V_\mu^{(+)} &=& -  \oint \oint_{|\lambda_1|>|\mu
|>|\lambda_2|}
{}~~{\cal I}
\nonumber
\end{eqnarray}
We reduce all the integrals to the contour
$|\mu |>|\lambda_1|>|\lambda_2|$.
Then, we have
\begin{eqnarray}
{V_\mu^{(-)}}^2 &=& - V_\mu^{(-)} V_\mu^{(+)} + {1\over 2}
\oint_{|\lambda
|>|\mu |} {d\lambda \over 2 i \pi \lambda } {\lambda -\mu \over \lambda
+ \mu
}  \norm V(\lambda ) V(\mu ) \norm \nonumber \\
 V_\mu^{(-)} V_\mu^{(+)}&=& -{ V_\mu^{(+)}}^2 -  {1\over 2}
 \oint_{|\lambda
 |<|\mu |} {d\lambda \over 2 i \pi \lambda } {\lambda -\mu \over
 \lambda + \mu
 }  \norm V(\lambda ) V(\mu ) \norm \nonumber
 \end{eqnarray}
It follows that
\begin{eqnarray}
{V_\mu^{(+)}}^2 + 2 V_\mu^{(-)} V_\mu^{(+)} +  {V_\mu^{(-)}}^2 ={1\over
4}
\nonumber
\end{eqnarray}
Now, one has
\begin{eqnarray}
N(V_\mu^{(-)} + V_\mu^{(+)} )^{n+1} = V_\mu^{(-)}N(V_\mu^{(-)} +
V_\mu^{(+)}
)^n +N(V_\mu^{(-)} + V_\mu^{(+)} )^n V_\mu^{(+)} \nonumber
\end{eqnarray}
and therefore
\begin{eqnarray}
N(V_\mu^{(-)} + V_\mu^{(+)} )^n &=& 2^{-n}
{}~~~~~~~~~~~~~~~~~~~~~~n~~even
\nonumber \\
N(V_\mu^{(-)} + V_\mu^{(+)} )^n &=&2^{-n+1}(V_\mu^{(-)} + V_\mu^{(+)})
{}~~n~~odd
\nonumber
\end{eqnarray}
and the result follows.

\section{Appendix F: The matrices $R(j,k)$.}

Next, we prove eq.(\ref{Eg+}). By direct computation, we have
\begin{eqnarray}
e^{{1\over 2} h H } V(\mu ) e^{-{1\over 2} h H } =
{1\over 2} (\cosh h +1 ) V(\mu ) - {1\over 2} (\cosh h -1 )
V(-\mu )
+ {i\over \sqrt{2}} \sinh h~~ \mu {d \over d \mu } Z(\mu ) \nonumber
\end{eqnarray}
\begin{eqnarray}
e^{{1\over 2} h H } {i\over \sqrt{2}}\mu {d\over d \mu }Z(\mu)
e^{-{1\over 2} h H } =\cosh h  {i\over \sqrt{2}}\mu {d\over d \mu
}Z(\mu) + {1\over 2} \sinh h [ V(\mu ) - V(-\mu )] \nonumber
\end{eqnarray}
or
\begin{eqnarray}
 e^{{1\over 2} h_j H }\bra{E(\mu_k)} e^{-{1\over 2} h_j H } &=&
\bra{ E(\mu_k)} R_1 (j,k)
\nonumber
\end{eqnarray}
with
\begin{eqnarray}
 R_1 (j,k) =\pmatrix{1 & 0 & 0 & 0 \cr
 0 & \cosh^2 {h_j \over 2} & - \sinh^2  {h_j \over 2} & {1\over
 2}
 \sinh h_j \cr
 0 &  - \sinh^2  {h_j \over 2} & \cosh^2 {h_j \over 2} &  -
 {1\over 2}
 \sinh h_j \cr
 0 &  \sinh h_j & - \sinh h_j & \cosh h_j } \nonumber
\end{eqnarray}
 also
\begin{eqnarray}
ad~ V_{\mu_j}^{(+)} \cdot V(\mu_k) & =& {i\over \sqrt{2}}
{\mu_j - \mu_k \over \mu_j + \mu_k }~~ \mu_k {d \over d \mu_k }
Z(\mu_k)
- {\mu_j \mu_k \over (\mu_j + \mu_k )^2 } ~~~~~~~~~|\mu_j | >
|\mu_k |
\nonumber \\
ad~ V_{\mu_j}^{(+)}\cdot  \mu_k {d \over d \mu_k } Z(\mu_k)  & =&
 - {i\over \sqrt{2}}\left[
 {\mu_j + \mu_k \over \mu_j -\mu_k }V(\mu_k) -{\mu_j - \mu_k \over
 \mu_j
 +\mu_k }V(-\mu_k) \right]  ~~~~~~|\mu_j | > |\mu_k | \nonumber
\end{eqnarray}
This can be rewritten in a more compact notation
\begin{eqnarray}
ad~ V_{\mu_j}^{(+)}\cdot \bra{E(\mu_k)}=\bra{E(\mu_k)}
\pmatrix{0& -{\mu_j \mu_k \over (\mu_j + \mu_k )^2 } &{\mu_j \mu_k
\over (\mu_j - \mu_k )^2 }& 0 \cr
0 & 0 & 0 & {1\over 2}{\mu_j + \mu_k \over \mu_j -\mu_k } \cr
0 & 0 & 0 &- {1\over 2}{\mu_j - \mu_k \over \mu_j +\mu_k } \cr
0 & {\mu_j - \mu_k \over \mu_j +\mu_k } &- {\mu_j + \mu_k \over \mu_j
-\mu_k } & 0 }
\nonumber
\end{eqnarray}
By exponentiation, we get
\begin{eqnarray}
 e^{-f_j V^{(+)}_{\mu_j} }\bra{E(k)} e^{f_j V^{(+)}_{\mu_j} } &=& \bra{
 E(k)} R_2
(j,k)
\nonumber
\end{eqnarray}
with
\begin{eqnarray}
R_2 (j,k) = \pmatrix{ 1 & {1\over 4} (1-\beta_{jk}^2 ) \sinh f_j &
{1\over 4}
(1-\beta_{jk}^{-2} ) \sinh f_j & {1\over 2} ( \beta_{jk}
-\beta_{jk}^{-1} )
\sinh^2 {f_j\over 2} \cr
0 & \cosh^2 {f_j\over 2} & - \beta_{jk}^{-2} \sinh^2 {f_j\over 2} & -
{1\over
2} \beta_{jk}^{-1} \sinh f_j \cr
0 & -\beta_{jk}^2 \sinh^2 {f_j\over 2} & \cosh^2 {f_j\over 2} & {1\over
2}
\beta_{jk} \sinh f_j \cr
0 & -\beta_{jk} \sinh f_j &\beta_{jk}^{-1} \sinh f_j & \cosh f_j }
\nonumber
\end{eqnarray}
{}From the definition of $g_+(k)$ in eq.(\ref{g+k}), we immediately get
eq.(\ref{Eg+}) with
\begin{eqnarray}
R(j,k)= R_2(j,k) R_1(j,k) \nonumber
\end{eqnarray}

\section{Appendix G.}

Finally, we prove eq.(\ref{nso01})
We notice that we can write
\begin{eqnarray}
R(j,k) =\pmatrix{1 & v(j,k) \cr 0 & \widetilde{R}(j,k) }
\nonumber
\end{eqnarray}
where $v(j,k)$ is a line vector which we will write for convenience as
$v(j,k) = \widetilde{v}
(j,k) \widetilde{R} (j,k) $ with
\begin{eqnarray}
\widetilde{v}(j,k)= \left( {1\over 4} (1-\beta_{jk}^2 ) \sinh f_j ~,~
{1\over
4}
(1-\beta_{jk}^{-2} ) \sinh f_j ~,~- {1\over 2} ( \beta_{jk} -\beta_{jk
}^{-1} )
\sinh^2 {f_j\over 2} \right) \label{vtilde}
\end{eqnarray}
and $\widetilde{R}$ is an explicitly known 3x3 matrix. With these
notations,
eq.(\ref{nso01}) is equivalent to the following set of two equations
\begin{eqnarray}
\widetilde{R}(1,k) \widetilde{R}(2,k) \cdots \widetilde{R}(k-1 ,k )
\pmatrix{1 \cr 0 \cr 0} ={\Delta_k X_k \over
\sinh {f_k \over 2}}\pmatrix{1 \cr 0 \cr 0 }
\label{3x3}
\end{eqnarray}
and
\begin{eqnarray}
\cosh {f_k \over 2} +2 \sinh {f_k \over 2}\sum_{j=1}^{k-1}
\widetilde{v}(j,k)\widetilde{R}(j,k) \widetilde{R}(j+1,k)
\cdots  \widetilde{R}(k-1,k) \pmatrix{1 \cr 0 \cr 0 } = \Delta_k
\label{nso3}
\end{eqnarray}
Eq.(\ref{3x3}) is an eigenvector problem. To show its consistency, we
remark that each $\widetilde{R}(j,k)$ is a matrix of the form
\begin{eqnarray}
\widetilde{R} =\pmatrix{ a^2 & -b^2 & ab \cr -c^2 & d^2 & - cd \cr
2 ac & -2 bd & 1 + 2 bc } \nonumber
\end{eqnarray}
This form and the condition
$ ad-bc =1$ are preserved by product. On the coefficients $a,b,c,d$ the
product
rule is just the
matrix product for the 2x2 matrices
\begin{eqnarray}
r(j,k) = \pmatrix{a(j,k) & b(j,k) \cr c(j,k) & d(j,k) } \nonumber
\end{eqnarray}
we have
\begin{eqnarray}
r(j,k) =\pmatrix{ \cosh {f_j \over 2} & - \beta_{jk}^{-1} \sinh
{f_j\over 2} \cr
- \beta_{jk} \sinh {f_j\over 2} &  \cosh {f_j \over 2} }
\pmatrix{ \cosh {h_j \over 2} &  \sinh {h_j \over 2} \cr
\sinh {h_j \over 2} &  \cosh {h_j \over 2} } \nonumber
\end{eqnarray}
Then eq.(\ref{3x3}) is equivalent to
\begin{eqnarray}
r(1,k) r(2,k) \cdots r(k-1,k) \pmatrix{1 \cr 0} =\alpha_k  \pmatrix{1
\cr 0}
\label{nso2}
\end{eqnarray}
with
\begin{eqnarray}
\alpha_k^2 = {\Delta_k X_k \over  \sinh {f_k\over 2}}
\label{eigenvalue}
\end{eqnarray}
 Eq.(\ref{nso2}) reduces to the single equation
\begin{eqnarray}
[r(1,k) r(2,k) \cdots r(k-1,k)]_{21} =0 \nonumber
\end{eqnarray}
This equation determines recursively $\tanh (h_{k-1}/2)$.
Therefore, for such $h_k$, eq.(\ref{3x3}) is consistent.
The eigenvalue is
\begin{eqnarray}
\alpha_k =[r(1,k) r(2,k) \cdots r(k-1,k)]_{11} \nonumber
\end{eqnarray}
It depends only on the fields $f_j~,~j\leq k-1 $. Let $\gamma_k$ be the
C-number
defined by
\begin{eqnarray}
\gamma_k = \sum_{j=1}^{k-1}\widetilde{v}(j,k)\widetilde{R}(j,k)
\widetilde{R}(j+1,k)
\cdots  \widetilde{R}(k-1,k) \pmatrix{1 \cr 0 \cr 0 } \nonumber
\end{eqnarray}
Again, this depends only on the fields $f_j~~j\leq k-1 $. Then
eq.(\ref{nso3})
gives
\begin{eqnarray}
\tanh {f_k \over 2} = {X_k \over \alpha_k^2 - 2 \gamma_k X_k }
\label{fk}
\end{eqnarray}
and this determines $f_k$ in terms of the $f_j,~~j\leq k-1$ and the
extra
variable $X_k$.

Consistency of the whole construction ensures that the fields $h_k$ and
$f_k$ so determined are the same as those of eq.(\ref{backlund1}-
\ref{backlund4}).

Let us do it for the two soliton case.  The condition $r(1,2)|_{21}=0$
gives
\begin{eqnarray}
\tanh {h_1\over 2} = \beta_{12} \tanh {f_1 \over 2} \nonumber
\end{eqnarray}
We recognize the `tangent rule'. Then we calculate
\begin{eqnarray}
\alpha_2 &=& {\cosh {h_1 \over 2} \over \cosh {f_1 \over 2} } \nonumber
= \sqrt{ 1-X_1^2 \over 1- \beta_{12}^2 X_1^2} \\
\gamma_2 &=& {1\over 4} (1-\beta_{12}^2) \sinh f_1  {\cosh^2 {h_1 \over
2}
\over \cosh^2 {f_1 \over 2} }
= {1\over 2} (1-\beta_{12}^2) {X_1 \over  1- \beta_{12}^2 X_1^2}
\nonumber
\end{eqnarray}
{}From eq.(\ref{fk}), we get
\begin{eqnarray}
e^{f_2} = {1-X_1 \over 1+X_1}\cdot
{1+X_1 +X_2 + \beta_{12}^2 X_1 X_2 \over 1-X_1 -X_2 + \beta_{12}^2 X_1
X_2} ={\tau_-(1) \over \tau_+(1)}\cdot{\tau_+(2) \over \tau_-(2)}
\nonumber
\end{eqnarray}
Finally, we can compute
\begin{eqnarray}
\Delta_2 = {\tau_0 \over \sqrt{ \tau_+(2) \tau_-(2) }}
\nonumber
\end{eqnarray}
as it should be.

\end{document}